\documentclass[letterpaper]{article}
\usepackage{arxivtemplate}
\usepackage{times}
\usepackage{helvet}
\usepackage{courier}
\usepackage[hyphens]{url}
\usepackage{graphicx}
\usepackage{natbib}
\usepackage{caption}
\usepackage{algorithm}
\usepackage{algorithmic}
\usepackage{amsmath}
\usepackage{amsfonts}
\usepackage{booktabs}
\usepackage{enumitem}
\usepackage{multirow}
\usepackage{caption}
\usepackage{subcaption}
\usepackage{newfloat}
\usepackage{listings}
\DeclareCaptionStyle{ruled}{labelfont=normalfont,labelsep=colon,strut=off}
\floatstyle{ruled}
\newfloat{listing}{tb}{lst}{}
\floatname{listing}{Listing}
\newcommand{\customsmall}{\fontsize{9}{8.9}\selectfont}

\title{Data-Efficient Neural Training with Dynamic Connectomes}
\author {
    Yutong Wu\textsuperscript{\rm 1},
    Peilin He\textsuperscript{\rm 2},
    Tananun Songdechakraiwut\textsuperscript{\rm 1}
}
\affiliations {
    \textsuperscript{\rm 1}Department of Computer Science, Duke University\\
    \textsuperscript{\rm 2}Division of Natural and Applied Sciences, Duke Kunshan University
}

\begin{document}

\maketitle

\begin{abstract}
The study of dynamic functional connectomes has provided valuable insights into how patterns of brain activity change over time. Neural networks process information through artificial neurons, conceptually inspired by patterns of activation in the brain. However, their hierarchical structure and high-dimensional parameter space pose challenges for understanding and controlling training dynamics. In this study, we introduce a novel approach to characterize training dynamics in neural networks by representing evolving neural activations as functional connectomes and extracting dynamic signatures of activity throughout training. Our results show that these signatures effectively capture key transitions in the functional organization of the network. Building on this analysis, we propose the use of a time series of functional connectomes as an intrinsic indicator of learning progress, enabling a principled early stopping criterion. Our framework performs robustly across benchmarks and provides new insights into neural network training dynamics.
\end{abstract}

\section{Introduction}

Deep neural networks have achieved remarkable success across a broad range of applications, from image recognition and language modeling to scientific discovery~\citep{he2016resnet, vaswani2017attention}. Despite this progress, the process of training deep networks remains challenging, with issues such as overfitting, instability, and suboptimal convergence often limiting practical performance. Among strategies for addressing these issues, early stopping is widely used to prevent overfitting by terminating training when model performance on a held-out validation set ceases to improve~\citep{prechelt1998automatic, goodfellow2016deep}.

While effective, standard early stopping techniques rely on partitioning data into separate training and validation sets. This approach poses difficulties in domains where labeled data is scarce, as holding out a validation set reduces the amount of data available for learning. Furthermore, validation performance may not always provide the most sensitive or timely signal for halting training, particularly under distribution shifts or when models generalize poorly beyond the validation set~\citep{recht2019imagenet}. These challenges motivate the search for alternative, data-efficient criteria to guide early stopping.

A promising direction is to monitor the \emph{internal dynamics} of the neural network itself during training. Recent advances in topological data analysis provide powerful tools for quantifying the evolving structure of high-dimensional data and complex networks~\citep{edelsbrunner2022computational}. In particular, persistent homology has emerged as a rigorous framework for capturing multiscale topological features in neural activations and connectivity patterns~\citep{rieck2018neural, hofer2017deep, zhang2024functional, songdechakraiwut2025functional}. However, the potential of topological signatures for informing training decisions, such as early stopping, remains largely unexplored.

In this work, we propose a novel \emph{connectome-guided early stopping} framework that leverages persistent homology to analyze the dynamic evolution of a neural network's functional organization during training. Rather than relying on external validation data, our method constructs a sequence of \emph{dynamic functional connectomes} based on correlations in neuron or channel activations across the training set. By applying persistent homology to these evolving connectivity graphs, we obtain a topological time series that summarizes structural changes in the network over time.

We show that monitoring the convergence of this topological time series, quantified via Wasserstein distances between persistence diagrams \citep{skraba2020wasserstein}, provides a robust, data-efficient criterion for early stopping. Our experiments on image classification benchmarks demonstrate that this approach achieves competitive or superior performance to validation-based early stopping, particularly when training data is limited. We also conduct direct comparisons with functional persistence~\citep{zhang2024functional}, a recent state-of-the-art method for topological monitoring in neural networks, and show that our approach yields improved or comparable results in terms of both efficiency and predictive performance.

Our main contributions are as follows:
\begin{itemize}
    \item We develop a new neuroimaging-inspired framework for tracking dynamic changes in neural network functional connectivity during training.
    \item We propose a connectome-guided early stopping criterion that does not require a separate validation set, enabling data-efficient training.
    \item We empirically validate our approach on multiple datasets and architectures, demonstrating improved or comparable performance to both validation-based early stopping and state-of-the-art functional persistence.
\end{itemize}

By connecting topological data analysis, neuroimaging approaches, and deep learning optimization, our work opens new directions for model monitoring, diagnostics, and training strategies in neural networks.

\section{Methods}

\subsection{Functional Connectome} \label{sec:connectome}

Consider a training set $\mathcal{X} = \{x^{(1)}, x^{(2)}, \dots, x^{(N)}\}$, where each sample $x^{(i)}$ is input to a neural network. In fully connected layers, the $j$th neuron applies an affine transformation followed by a nonlinearity to each input $x^{(i)}$, resulting in an activation signal $s_{ij}$. Across the training set, the activations of the $j$th neuron form an activation vector $\mathbf{a}_j = [s_{1j}, s_{2j}, \dots, s_{Nj}]$.
To quantify the statistical dependency between two neurons, we compute the Pearson correlation coefficient between their activation vectors. For neurons $i$ and $j$, the Pearson correlation is defined as
$\rho_{ij} = \mathrm{Cov}(\mathbf{a}_i, \mathbf{a}_j) / \big(\sigma(\mathbf{a}_i)\sigma(\mathbf{a}_j)\big)$,
where $\mathrm{Cov}$ denotes covariance and $\sigma$ denotes standard deviation. This approach has been adopted in prior studies~\citep{songdechakraiwut2025functional} to characterize functional connectivity in feedforward artificial neural networks.

This construction extends naturally to convolutional neural networks (CNNs). For a convolutional layer, given an input $x^{(i)} \in \mathbb{R}^{C_\text{in} \times H \times W}$, the $j$th filter produces an activation map $\mathbf{z}_{ij} \in \mathbb{R}^{H' \times W'}$. In this work, we propose to define the activation vector for each channel $j \in \{1, \dots, C_\text{out}\}$ as the collection of activation maps across the training set:
\[
\mathbf{A}_j = [\mathbf{z}_{1j}, \mathbf{z}_{2j}, \dots, \mathbf{z}_{Nj}],
\]
where each $\mathbf{z}_{ij}$ is an activation map corresponding to sample $i$ and channel $j$.

To measure the similarity or dependency between two such activation vectors $\mathbf{A}_i$ and $\mathbf{A}_j$, we may compute the Pearson correlation coefficient by first mapping each activation map to a scalar using a reduction function $R: \mathbb{R}^{H' \times W'} \rightarrow \mathbb{R}$. Typical choices for $R$ include max pooling $R_{\mathrm{max}}(\mathbf{z}_{ij}) = \max_{u,v} (\mathbf{z}_{ij}[u, v])$, mean pooling $R_{\mathrm{mean}}(\mathbf{z}_{ij}) = \frac{1}{H'W'} \sum_{u=1}^{H'} \sum_{v=1}^{W'} \mathbf{z}_{ij}[u, v]$, or a global norm $R_{\mathrm{L2}}(\mathbf{z}_{ij}) = \frac{1}{H'W'} \sum_{u=1}^{H'} \sum_{v=1}^{W'} (\mathbf{z}_{ij}[u, v])^2$.
Applying $R$ to each activation map yields a scalar activation signal $s_{ij} = R(\mathbf{z}_{ij})$, and thus a reduced activation vector $\mathbf{a}_j = [s_{1j}, s_{2j}, \dots, s_{Nj}]$ for each channel $j$. The Pearson correlation can then be computed between these reduced activation vectors.

With these definitions, we construct the \emph{functional connectome} of a neural network as a weighted adjacency matrix $\mathbf{M} \in \mathbb{R}^{N \times N}$, where each off-diagonal entry $\mathbf{M}_{ij}$ (for $i \neq j$) is given by the absolute value of the Pearson correlation $|\rho_{ij}|$ between neurons or channels $i$ and $j$, and diagonal entries are set to zero to exclude self-loops.

\subsection{Dynamic Functional Connectome}

\begin{figure}[t]
\centering
\includegraphics[width=\columnwidth]{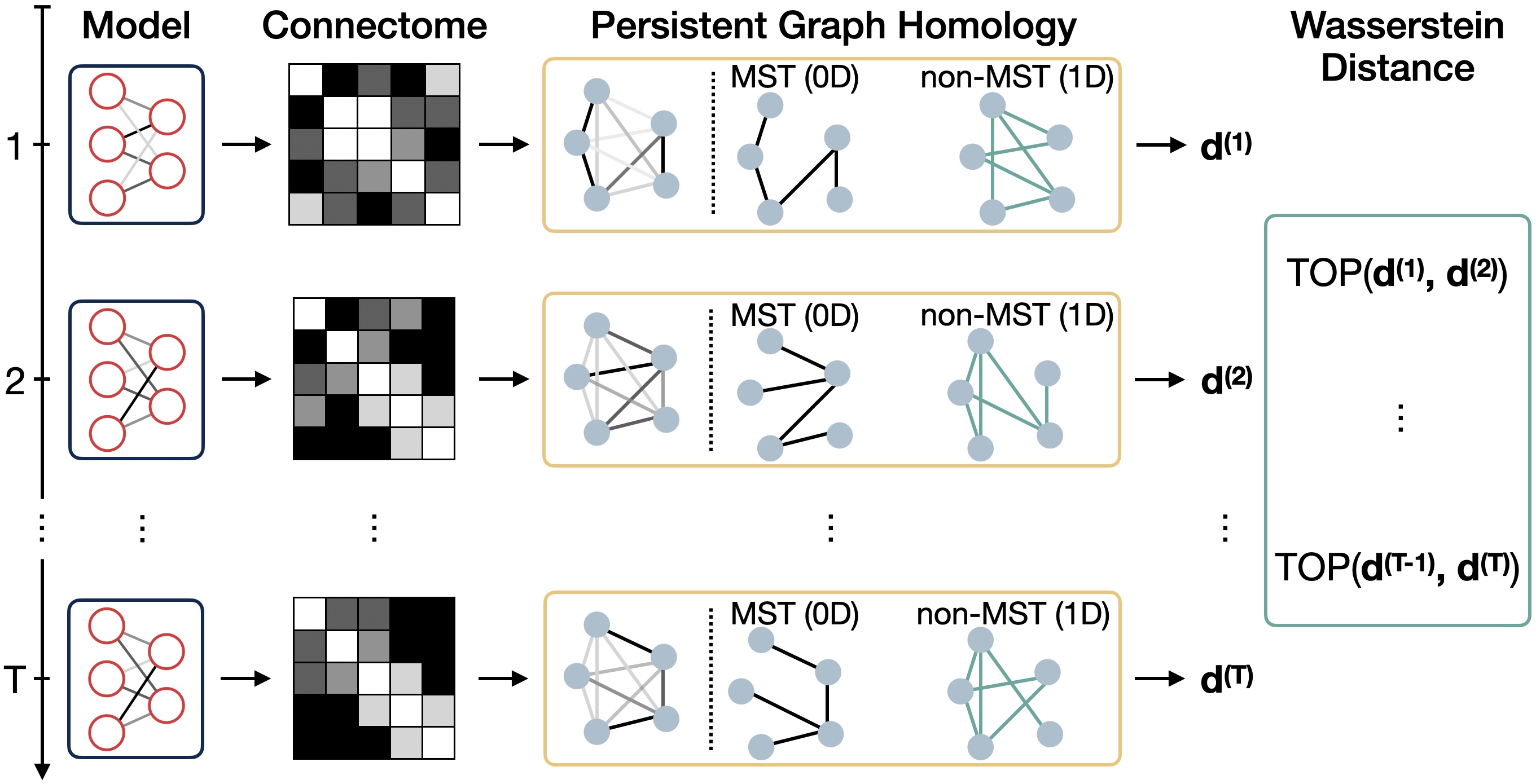} 
\caption{Schematic of constructing the topological time series for TOP. Model states and functional connectomes are recorded at each epoch. The Wasserstein distances between consecutive connectomes form the values of the time series.}
\label{fig:schematic}
\end{figure}

Dynamic functional connectivity (dFC) is widely used to characterize changing patterns of connectivity in complex networks. Originally developed for brain connectivity analysis  \cite{leonardi2015spurious}, dFC provides a framework for studying how functional relationships between network units evolve over time. In the context of artificial neural networks, the training process induces continuous changes in functional organization as the network adapts to data. By adopting the dFC perspective, we analyze how the functional connectome of a neural network evolves throughout training.
To analyze this dynamic organization, a functional connectome is constructed at the end of each training epoch, based on the activation patterns of all neurons or channels. As training progresses, a time series of connectomes is collected over $T$ epochs, denoted by $\mathcal{S} = (\mathbf{M}^{(1)}, \mathbf{M}^{(2)}, \dots, \mathbf{M}^{(T)})$ and termed the \textit{dynamic functional connectomes}.

To quantify the dynamically changing topology of these connectomes, a topological distance is computed between each pair of consecutive connectomes: $\text{dist}_k = \operatorname{dist}(\mathbf{M}^{(k)}, \mathbf{M}^{(k+1)})$. This yields a sequence of topological distance values, $\mathcal{F} = (\text{dist}_1, \text{dist}_2, \dots, \text{dist}_{T-1})$, termed the \textit{topological time series}. The topological time series is used for early stopping, as described in Section~\ref{sec:early_stopping}. Figure~\ref{fig:schematic} provides an overview of how topological changes are monitored during training.

\begin{algorithm}[t]
\customsmall
\caption{Online Connectome-Guided Early Stopping}
\label{alg:early-stopping}
\begin{algorithmic}[1]
\REQUIRE Training set $\mathcal{X}$; maximum epochs $T$; burn-in $b$; patience window $p$; threshold $\epsilon$
\ENSURE Stopping epoch $t^*$

\STATE $t^\star \leftarrow b$, \texttt{streak} $\leftarrow 0$ 

\FOR{$t = 1$ to $T$}
    \STATE $\mathbf{M}^{(t)}_{ij} \leftarrow \begin{cases}
        |\mathrm{corr}(\mathbf{a}_i^{(t)}, \mathbf{a}_j^{(t)})| & i \neq j \\
        0 & i = j
    \end{cases}\quad \forall\, i,j$
    
    \STATE $T = (V, E_T) \leftarrow$ MST of graph $\mathbf{M}^{(t)}$
    \STATE $E_{\mathrm{loop}} \leftarrow E \setminus E_T$
    \STATE $\mathbf{d}^{(t)} \leftarrow \mathrm{sort}\big(\{w(e) : e \in E_{\mathrm{loop}}\}\big)$

    \STATE \textit{// Or, compute $\mathrm{PD}^{(t)} \leftarrow \mathrm{PersistentHomology}(\mathbf{M}^{(t)})$}

    \IF{$t > 1$}
        \STATE $w_\mathrm{TOP}^{(t)} \leftarrow \mathrm{sum}(\mathrm{abs}(\mathbf{d}^{(t)} - \mathbf{d}^{(t-1)}))$
        \STATE \textit{// Or, compute $w_\mathrm{WD}^{(t)} \leftarrow \mathrm{WD}(\mathrm{PD}^{(t-1)}, \mathrm{PD}^{(t)})$}
    \ENDIF

    \IF{$t \ge b$}
        \IF{$w_{\mathrm{TOP}}^{(t^\star)} - w_{\mathrm{TOP}}^{(t)} > \epsilon$}
            \STATE \textit{// Or, use $w_{\mathrm{TOP}}^{(t^\star)} - w_\mathrm{WD}^{(t)} > \epsilon$}
            \STATE $t^\star \leftarrow t$, \texttt{streak} $\leftarrow 0$
        \ELSE
            \STATE \texttt{streak} $\leftarrow$ \texttt{streak} $+ 1$
        \ENDIF
        \IF{\texttt{streak} $= p$}
            \STATE $t^* \leftarrow t$
            \STATE \textbf{break}
        \ENDIF
    \ENDIF
\ENDFOR

\STATE \textbf{Return} $t^*$
\end{algorithmic}
\end{algorithm}

\subsection{Persistence-Based Topological Time Series} \label{sec:ph}

To rigorously and efficiently capture the evolving topology summarized by the topological time series, we turn to \emph{Persistent Graph Homology}, which provides topological invariants of neural networks represented as graphs \cite{songdechakraiwut2023topological}.
We start with a symmetric adjacency matrix $\mathbf{M}$, defined in Section~\ref{sec:connectome}. From $\mathbf{M}$, we define a weighted undirected graph $G = (V, E, w)$, where $V$ is the set of neurons, $E$ is the set of edges, and $w: E \rightarrow \mathbb{R}_{\geq 0}$ assigns weights based on pairwise similarity. Typically, we set $w(i, j) = \mathbf{M}_{ij}$. Then, we extract the Maximum Spanning Tree (MST) from $G$. We denote it as $T = (V, E_T, w_T)$, where $E_T \subseteq E$. The MST keeps the strongest connections while ensuring global connectivity. Each edge $e \in E_T$ connects two previously separate components. The edge weights $w(e)$ can be interpreted as 0-dimensional persistence values, reflecting the merging of connected components during filtration. 

The remaining edges, denoted by $E_{\mathrm{loop}} := E \setminus E_T$, form cycles within the graph, representing 1-dimensional topological features (i.e., loops) \cite{songdechakraiwut2021topological}. Each loop edge has an associated \emph{death time}, defined as its weight $w(e)$, which indicates the scale at which the loop is absorbed into the network structure. We collect these death times and sort them in ascending order. This gives us a \emph{persistence vector} at each training step,
$\mathbf{d}^{(t)} = [d_1^{(t)}, d_2^{(t)}, \dots, d_k^{(t)}]$, with $d_1^{(t)} \leq \cdots \leq d_k^{(t)}$,
which gives a compact and interpretable summary focused on loops \cite{songdechakraiwut2023wasserstein}. This makes it well-suited for tracking how networks reorganize during learning.

Within our framework, we also use standard persistent homology. We apply a Vietoris–Rips filtration at each training step $t$. We begin by computing a dissimilarity matrix $\mathbf{D}^{(t)} \in \mathbb{R}^{M \times M}$, where $D_{ij}^{(t)} = 1 - |\rho_{ij}|$ and $\rho_{ij}$ is the Pearson correlation between activations of neurons $i$ and $j$. We then apply persistent homology to $\mathbf{D}^{(t)}$ to obtain the two-dimensional persistence diagram $\mathrm{PD}^{(t)}$, where each point $(b_i^{(t)}, d_i^{(t)})$ denotes the birth and death of a loop. For implementation details, see Appendix.

To track the evolution of topological features during training, we compute the $p$-Wasserstein distance between successive persistence diagrams. This metric is known to be stable under perturbations to the input data \cite{skraba2020wasserstein}, which justifies its use in noisy training environments. Such stability also enables statistical analysis of topological time series, including the construction of confidence intervals (CIs) based on empirical variability across independently trained models. Formally, the \emph{$p$-Wasserstein distance} compares persistence diagrams or persistence vectors:
$W_p(X, Y) = \left( \inf_{\gamma \in \Gamma(X, Y)} \sum_{(x, y) \in \gamma} \|x - y\|^p \right)^{1/p}$,
where $X$ and $Y$ are two multisets of points. The set $\Gamma(X, Y)$ includes all valid matchings, including optional diagonal projections for unmatched features.

In our experiments, for Persistent Graph Homology, we use $p = 1$ and compute the Wasserstein distance between persistence vectors, which we refer to as $\mathrm{TOP}$:
\[
\mathrm{TOP}(\mathbf{d}^{(t)}, \mathbf{d}^{(t+1)}) = \sum_{j=1}^{k} \left| d_j^{(t)} - d_j^{(t+1)} \right|.
\]
For standard Persistent Homology, we use $p = 2$ and compute the Wasserstein distance between persistence diagrams, which we refer to as $\mathrm{WD}$:
  \begin{align*}
\mathrm{WD}(&\mathrm{PD}^{(t)}, \mathrm{PD}^{(t+1)}) \\
&= \left( \inf_{\gamma} \sum_{i} \left\| (b_i^{(t)}, d_i^{(t)}) - \gamma(b_i^{(t+1)}, d_i^{(t+1)}) \right\|_2^2 \right)^{1/2}.
  \end{align*}

By computing $p$-Wasserstein distances between each training step, we form a topological time series. This sequence quantifies how the network’s structure changes over time. It supports downstream tasks such as change-point detection or computing descriptive statistics like mean and variance.

In addition, we can also utilize bottleneck distance (BD), heat kernel (HK) \cite{reininghaus2015stable}, and sliced Wasserstein kernel
(SWK) \cite{carriere2017sliced}. Details are available in the supplementary material.

\begin{figure*}[t]
  \centering
  \includegraphics[width=\textwidth]{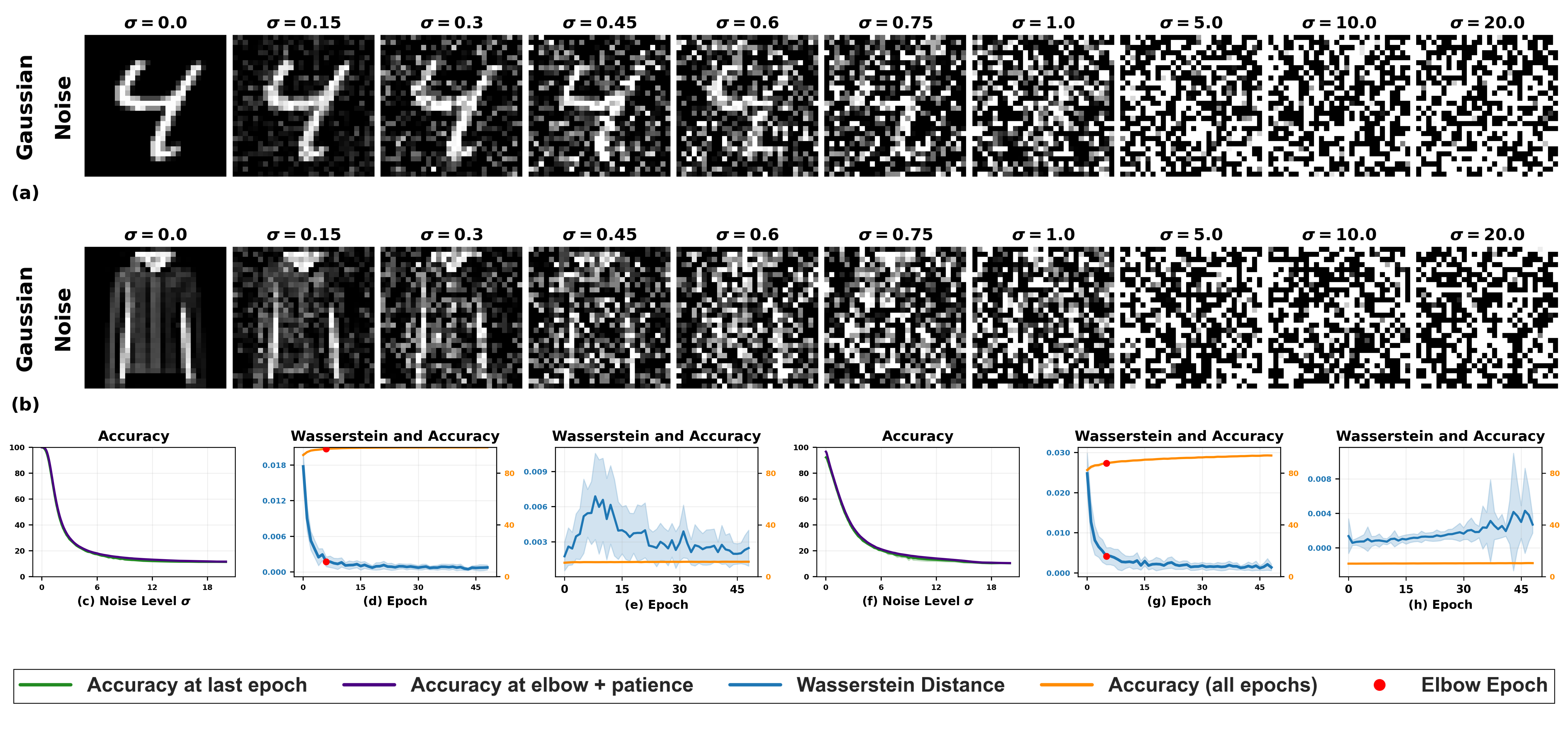}
\caption{
(a)-(b): Example MNIST (a) and Fashion-MNIST (b) images with increasing Gaussian noise levels ($\sigma$ indicated above).  
(c), (f): Test accuracy as a function of noise level $\sigma$, comparing accuracy at the last epoch and at (elbow + patience = 10) epoch for MNIST (c) and Fashion-MNIST (f).  
(d), (g): Joint evolution of test accuracy and 1D Wasserstein distance (TOP) across training epochs for low noise ($\sigma = 0.15$) in MNIST (d) and Fashion-MNIST (g).  
(e), (h): Same as (d), (g) for high noise ($\sigma = 20.0$).  
}
  \label{fig:study1}
\end{figure*}

\subsection{Connectome-Guided Early Stopping} \label{sec:early_stopping}
Traditional early stopping methods require a portion of the training data to be reserved as a validation set. This can be especially problematic when working with small datasets, since it reduces the amount of data available for training the model itself \cite{goodfellow2016deep}. To address this limitation, we propose a new connectome-guided early stopping strategy. Our method does not require a separate validation set. Instead, it uses the model’s topological behavior during training. The key idea is to track how the network’s internal structure evolves.

Specifically, we measure the $p$-Wasserstein distance between topological features of consecutive training epochs. After an initial burn-in phase in which the model is allowed to converge, we compute this distance at each epoch and observe its changes over time. If the distance ceases to decrease significantly for a specified number of epochs, we conclude that the model has converged and stop training.

Let $\{ w^{(t)} \}_{t=1}^{T}$ be the sequence of Wasserstein distances. Recall $T$ is the total number of epochs. Let $b$ be the burn-in rate and $p$ denotes the patience window. We define the stopping time $t^*$ as:
\[
t^*=\min\Bigl\{\,t\ge b+p \;\big|\, w^{(t-p)}-w^{(t)}\le\epsilon \Bigr\},
\]
where the condition must hold for $p$ consecutive epochs. Here, $\epsilon$ is a user-defined threshold, which specifies the minimum amount of change we consider meaningful. This allows us to stop training once topological changes have converged, offering a data-efficient alternative to traditional early stopping (see Algorithm~\ref{alg:early-stopping}).

\section{Experiments}

\paragraph{Datasets.}
We conducted experiments over four image datasets: MNIST \cite{lecun1998gradient}, Fashion-MNIST \cite{xiao2017fashion}, CIFAR-10, and CIFAR-100 \cite{krizhevsky2009learning}. MNIST consists of grayscale images of handwritten digits; Fashion-MNIST consists of grayscale images of clothing and fashion items; CIFAR-10 and CIFAR-100 consist of colored images of animals and everyday objects. MNIST, Fashion-MNIST, and CIFAR-10 each consist of 10 predefined classes, while CIFAR-100 consists of 100 predefined classes.

\paragraph{Architecture and optimization.}
We used two neural network architectures matched to dataset complexity: a 2-layer MLP for MNIST and Fashion-MNIST, and three VGG blocks followed by a 2-layer MLP for CIFAR-10/100~\cite{simonyan2014very}. Stochastic gradient descent was used for MNIST and Fashion-MNIST, and the Adam optimizer for CIFAR-10/100~\cite{KingmaB14adam}. Model configurations are shown in Table~1, and hyperparameter details are provided in Appendix.

\begin{table}
\centering
\customsmall
\setlength{\tabcolsep}{3mm}
\begin{tabular}{lccc}
\toprule
Layer    & \textbf{MNIST/F-MNIST} & \textbf{CIFAR-10/CIFAR-100}    \\
\midrule
VGG1   & -- & [3,\,32],\,[32,\,32]  \\
VGG2   & --  & [32,\,64],\,[64,\,64]\\
VGG3   & --  & [64,\,128],\,[128,\,128]\\
\midrule
Flatten  & [784]   & [2048]                          \\
\midrule
FC1      & [784,\,300] & [2048,\,300]                    \\
FC2      & [300,\,100] & [300,\,100]                     \\
FC3      & [100,\,10] & [100,\,10/100]                      \\
\midrule
Output   & 10            & 10/100               \\
\bottomrule
\end{tabular}
\caption{Model architectures for MNIST, Fashion-MNIST, CIFAR-10, and CIFAR-100. Each layer is shown as [\emph{in-channels},\,\emph{out-channels}]}
\end{table}

\begin{table*}
  \centering
  \customsmall
  \setlength{\tabcolsep}{2.8mm}
  
  \begin{subtable}[t]{\textwidth}
    \centering
    \begin{tabular}{l l l l l l}
    \toprule
      Dataset & Method & $\Delta$ Epoch (VL) & $\Delta$ Acc.\,(VL)\% &
                       $\Delta$ Epoch (FP) & $\Delta$ Acc.\,(FP)\%\\
      \midrule
        \multirow{5}{*}{\textbf{MNIST}}
        & TOP & \textbf{-2.953\,(-3.155, -2.750)} & \textbf{1.571\,(1.474, 1.668)} & \textbf{-9.032\,(-9.424, -8.641)} & -1.030\,(-1.132, -0.927)\\
        & WD  & 21.983\,(21.385, 22.581) & \textbf{2.702\,(2.682, 2.722)} & 15.903\,(15.511, 16.295) & \textbf{0.102\,(0.086, 0.118)}\\
        & SWK & 21.330\,(20.731, 21.929) & \textbf{2.717\,(2.700, 2.734)} & 15.250\,(14.860, 15.640) & \textbf{0.117\,(0.105, 0.128)}\\
        & BD  &  5.554\,(5.201, 5.908) & \textbf{2.532\,(2.502, 2.562)} & \textbf{-0.526\,(-0.695, -0.356)} & -0.068\,(-0.098, -0.039)\\
        & HK  &  3.328\,(3.177, 3.478) & \textbf{2.542\,(2.530, 2.555)} & \textbf{-2.752\,(-2.870, -2.634)} & -0.058\,(-0.068, -0.048)\\
        \midrule
        \multirow{5}{*}{\shortstack{\textbf{Fashion-}\\\textbf{MNIST}}}
        & TOP & \textbf{-3.264\,(-3.419, -3.110)} & \textbf{0.696\,(0.612, 0.781)} & \textbf{-17.174\,(-17.804, -16.543)} & -1.352\,(-1.444, -1.260)\\
        & WD  & {-0.174\,(-0.447, 0.099)} & \textbf{0.817\,(0.730, 0.904)} & \textbf{-14.083\,(-14.766, -13.399)} & -1.231\,(-1.324, -1.137)\\
        & SWK &  1.298\,(1.033, 1.563) & \textbf{1.071\,(1.007, 1.135)} & \textbf{-12.611\,(-13.247, -11.975)} & -0.976\,(-1.048, -0.904)\\
        & BD  & \textbf{-0.237\,(-0.406, -0.067)} & \textbf{1.061\,(1.000, 1.123)} & \textbf{-14.146\,(-14.703, -13.589)} & -0.988\,(-1.056, -0.920)\\
        & HK  &  4.264\,(4.040, 4.488) & \textbf{1.514\,(1.488, 1.538)} & \textbf{-9.645\,(-10.088, -9.203)} & -0.534\,(-0.564, -0.504)\\
        \bottomrule
    \end{tabular}
    \caption{5\% Train / 5\% Validation}
    \label{table:mnist5to5}
  \end{subtable}

\vspace{1ex}

  \begin{subtable}[t]{\textwidth}
    \centering
    \begin{tabular}{l l l l l l}
    \toprule
        Dataset & Method & $\Delta$ Epoch (VL) & $\Delta$ Acc. (VL) (\%) &
                           $\Delta$ Epoch (FP) & $\Delta$ Acc. (FP) (\%)\\
        \midrule
        \multirow{5}{*}{\textbf{MNIST}}
        & TOP & \textbf{-3.010\,(-3.227, -2.792)} & -0.729\,(-0.829, -0.628) & \textbf{-8.894\,(-9.281, -8.508)} & -1.028\,(-1.131, -0.926)\\
        & WD  & 21.925\,(21.332, 22.518) & \textbf{0.402\,(0.384, 0.420)} & 16.041\,(15.643, 16.438) & \textbf{0.102\,(0.086, 0.118)}\\
        & SWK & 21.289\,(20.694, 21.884) & \textbf{0.418\,(0.404, 0.432)} & 15.405\,(15.008, 15.801) & \textbf{0.118\,(0.107, 0.130)}\\
        & BD  &  5.502\,(5.154, 5.850) & \textbf{0.232\,(0.202, 0.263)} & \textbf{-0.383\,(-0.555, -0.211)} & -0.067\,(-0.097, -0.038)\\
        & HK  &  3.287\,(3.138, 3.437) & \textbf{0.243\,(0.232, 0.255)} & \textbf{-2.598\,(-2.710, -2.485)} & -0.056\,(-0.066, -0.046)\\
        \midrule
        \multirow{5}{*}{\shortstack{\textbf{Fashion-}\\\textbf{MNIST}}}
        & TOP & \textbf{-3.794\,(-3.956, -3.631)} & -1.147\,(-1.233, -1.060) & \textbf{-17.243\,(-17.872, -16.615)} & -1.342\,(-1.434, -1.250)\\
        & WD  & \textbf{-0.732\,(-1.009, -0.454)} & -1.002\,(-1.093, -0.911) & \textbf{-14.181\,(-14.862, -13.501)} & -1.197\,(-1.294, -1.101)\\
        & SWK &  0.644\,(0.377, 0.911) & -0.752\,(-0.818, -0.686) & \textbf{-12.806\,(-13.442, -12.169)} & -0.947\,(-1.019, -0.875)\\
        & BD  & \textbf{-0.700\,(-0.871, -0.529)} & -0.793\,(-0.856, -0.731) & \textbf{-14.150\,(-14.701, -13.598)} & -0.988\,(-1.056, -0.921)\\
        & HK  &  3.917\,(3.709, 4.124) & -0.237\,(-0.263, -0.211) & \textbf{-9.533\,(-9.957, -9.109)} & -0.432\,(-0.461, -0.403)\\
        \bottomrule
    \end{tabular}
    \caption{9\% Train / 1\% Validation}
    \label{table:mnist9to1}
\end{subtable}

   \caption{Differences in training performance between topological time series and baselines on MNIST and Fashion-MNIST. Results are reported in sample mean and 95\% CI. Earlier stops (negative gap) and higher test accuracy (positive gap) are \textbf{bolded}.}
   \label{table:mnist}
\end{table*}

\subsection{Study 1: Topological Convergence Study}

We evaluate functional connectome convergence under varying levels of distribution shift, where networks are trained on a shifted training set $\tilde{\mathcal{X}}$ and tested on original $\mathcal{X}_{\text{test}}$.
To simulate distribution shift, we add Gaussian noise to the training set: for each $x_{\text{train}}^{(i)} \in \mathcal{X}$, we define $\tilde{x}^{(i)} = x_{\text{train}}^{(i)} + \epsilon^{(i)}$, where $\epsilon^{(i)} \sim \mathcal{N}(0, \sigma^2)$. For each noise level $\sigma$, we independently train 20 neural networks and, for each, analyze the evolution and convergence of functional connectomes constructed from $\tilde{\mathcal{X}}$.

To detect convergence, we apply the elbow method to the sequence of $\mathrm{TOP}$ distances between persistence vectors from consecutive epochs. The elbow method identifies the convergence epoch $t^*$, which we interpret as the point where the functional connectome has converged. We then compare the distributions of $\mathrm{TOP}$ distance and test accuracy at $t^*$ across noise levels and compare them to the test accuracy at the maximum epoch.

To illustrate, Figure~\ref{fig:study1} shows two scenarios: under low noise ($\sigma = 0.15$), connectome convergence aligns with accuracy saturation, indicating successful learning. Under high noise ($\sigma = 20.0$), convergence fails and accuracy remains low. These results demonstrate that convergence of the functional connectome, measured via $\mathrm{TOP}$ distance, can reliably signal learning success and serve as an early stopping indicator under distribution shift.

\subsection{Study 2: Model Training Dynamics}

\begin{figure*}[t]
\centering
\includegraphics[width=\textwidth]{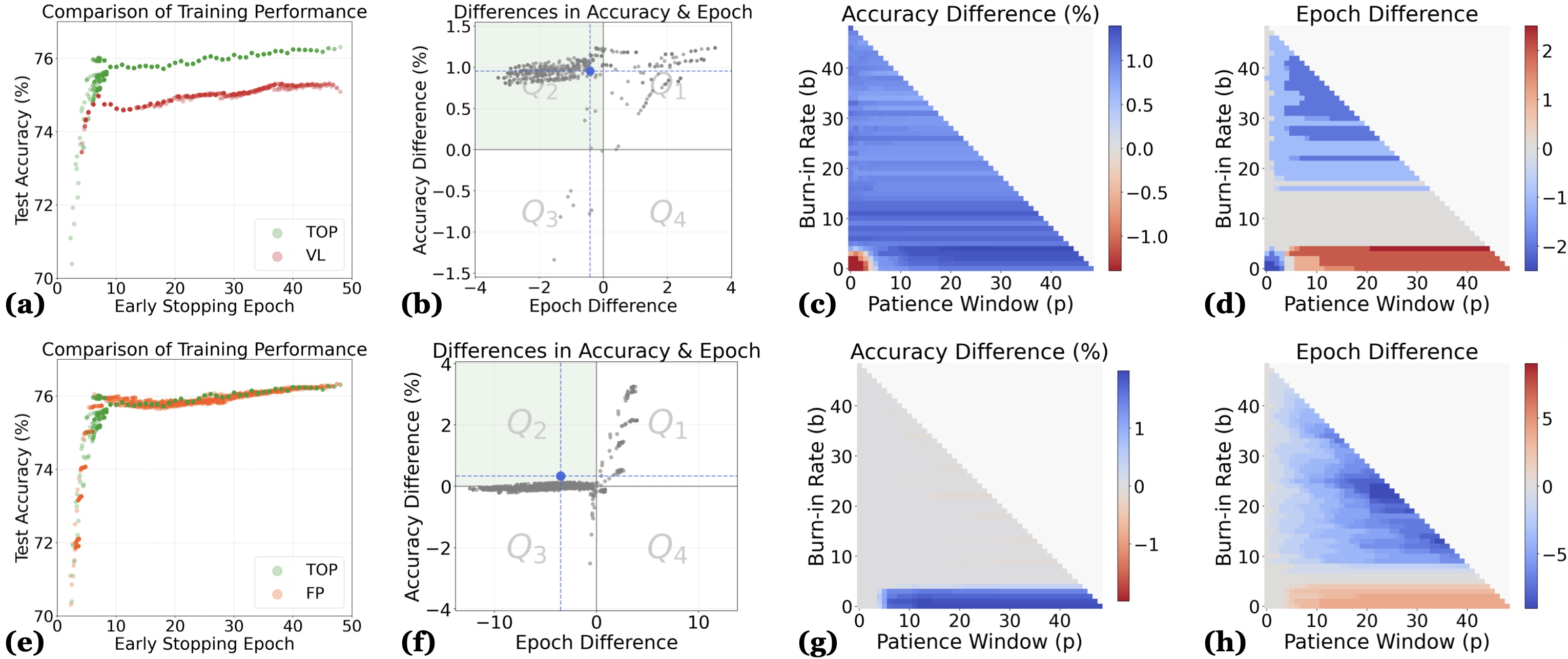} 
\caption{The training performance of TOP versus VL and FP on CIFAR-10. (a)-(d) compare TOP with VL. (e)-(h) compare TOP with FP. Scatter plots (a), (b), (e), (f) plot the raw early-stopping epoch and test accuracy and their paired differences; each data point corresponds to an early-stopping setting specified by burn-in rate $b$ and patience window $p$. The blue dots in (b) and (f) mark the mean difference. Heatmaps (c), (d), (g), (h) illustrate the difference in accuracy and epoch across $(b,p)$ settings. Blue denotes higher test accuracy or earlier stop in training for TOP relative to the baseline. }
\label{fig:study2}
\end{figure*}

We investigate the training dynamics and evaluate different topological criteria for early stopping based on their impact on training performance. Using these criteria, we apply our early stopping method and monitor two key metrics: the number of epochs before stopping and the test accuracy at the stopping point. We then compare performance between the different topological distance metrics and established baselines to assess their effectiveness.

The early stopping criterion is parameterized by the burn-in rate $b$ and the patience window $p$. For each configuration $(b, p)$, early stopping is triggered at epoch $t_{(b,p)}^*$, and the corresponding test accuracy is recorded as $a_{(b,p)}^*$. For a training process of $T$ epochs, we define two $T \times T$ lower triangular matrices to summarize performance across parameter settings: the epoch matrix $\mathbf{B}$, where $\mathbf{B}_{ij} = t_{(i,j)}^*$, and the accuracy matrix $\mathbf{A}$, where $\mathbf{A}_{ij} = a_{(i,j)}^*$. Here, $i$ and $j$ index the burn-in and patience parameters, respectively, under the constraint $b + p \leq T$.

For evaluation, we consider several topological criteria based on our proposed connectome-guided framework: TOP, WD, BD, HK, and SWK, as described in Section~\ref{sec:ph}. These are compared against the established baselines of validation loss~(VL) and state-of-the-art functional persistence~(FP).

For each criterion and baseline, we perform 100 independent training trials. In each trial, we compute the $T \times T$ epoch and accuracy matrices, yielding a collection of 100 epoch matrices $\{\mathbf{B}^{(k)}\}_{k=1}^{100}$ and 100 accuracy matrices $\{\mathbf{A}^{(k)}\}_{k=1}^{100}$ for each setting. To compare performance, we calculate the paired difference in stopping epoch and test accuracy for each trial: 
\[
\mathbf{A}_{\text{diff}}^{(k)} = \mathbf{A}_{\text{method}}^{(k)} - \mathbf{A}_{\text{baseline}}^{(k)}, \quad
\mathbf{B}_{\text{diff}}^{(k)} = \mathbf{B}_{\text{method}}^{(k)} - \mathbf{B}_{\text{baseline}}^{(k)},
\]
where $\mathbf{A}_{\text{method}}^{(k)}$ and $\mathbf{B}_{\text{method}}^{(k)}$ denote the accuracy and epoch matrices, respectively, for a given topological criterion, and $\mathbf{A}_{\text{baseline}}^{(k)}$ and $\mathbf{B}_{\text{baseline}}^{(k)}$ are the corresponding matrices for the baseline method. We then aggregate results by computing the sample mean and $95\%$ confidence interval for the difference in accuracy and epochs, elementwise across the $T \times T$ matrices over the 100 trials, for each parameter pair $(b, p)$. Figure \ref{fig:study2} illustrates the raw training performance of TOP and its paired difference to baselines on CIFAR-10.

To evaluate the robustness and generalization of each early stopping criterion, we conduct experiments across a diverse set of datasets and architecture combinations. These include scenarios with restricted training data, as well as settings where the dataset complexity presents significant challenges for the given model architecture. This experimental design allows us to systematically assess the performance of each early stopping method under a variety of realistic and demanding conditions.

\subsection*{Scenario 1: Data-Limited Regime}

To evaluate performance in data-scarce settings, we train all topological early stopping methods (TOP, WD, BD, HK, and SWK, as well as the state-of-the-art FP baseline) using only 10\% of the original training data, selected via stratified sampling. This same subset is used for both training and constructing the topological time series on MNIST and Fashion-MNIST.
For the VL baseline, we follow standard practice and split the 10\% subset into training and validation portions. We consider two splits: in the first, 9\% of the data is used for training and 1\% for validation; in the second, 5\% is used for training and 5\% for validation. This allows for a fair comparison under comparable data budgets.

Under the 5\%/5\% split, test accuracy measured at the maximum epoch is $92.80\% \pm 0.14\%$ for MNIST and $81.88\% \pm 0.65\%$ for Fashion-MNIST. Table~\ref{table:mnist5to5} summarizes the differences. On both datasets, TOP achieves earlier stopping ($-2.953$ and $-3.264$ epochs) and higher test accuracy (1.571\% and 0.696\%) compared to VL, with statistically significant improvements confirmed by 95\% confidence intervals well above zero for accuracy and well below zero for stopping epochs. Compared to the state-of-the-art FP, TOP generally stops earlier with a modest decrease in accuracy.
Under the 9\%/1\% split, test accuracy reaches $94.92\% \pm 0.11\%$ for MNIST and $83.92\% \pm 0.80\%$ for Fashion-MNIST at the maximum epoch. As shown in Table~\ref{table:mnist9to1}, similar trends to those observed under the 5\%/5\% split are observed.

\begin{table*}[ht]
  \centering
  \customsmall
  \setlength{\tabcolsep}{3.2mm}
  \begin{subtable}[t]{\textwidth}
    \centering
    \begin{tabular}{l l l l l l}
    \toprule
        Dataset & Method & $\Delta$ Epoch (VL) & $\Delta$ Acc. (VL) (\%) &
                           $\Delta$ Epoch (FP) & $\Delta$ Acc. (FP) (\%)\\
        \midrule
        \multirow{5}{*}{\textbf{CIFAR-10}}
        & TOP & \textbf{-0.413\,(-0.510, -0.316)} & \textbf{0.954\,(0.925, 0.983)} & \textbf{-3.522\,(-3.763, -3.281)} & \textbf{0.332\,(0.280, 0.383)}\\
        & WD  & 14.725\,(14.128, 15.323) & \textbf{1.242\,(1.216, 1.267)} & 11.616\,(11.029, 12.203) & \textbf{0.619\,(0.556, 0.683)}\\
        & SWK & 13.707\,(13.138, 14.275) & \textbf{1.244\,(1.221, 1.267)} & 10.597\,(10.043, 11.152) & \textbf{0.622\,(0.558, 0.686)}\\
        & BD  &  2.150\,(1.996, 2.304) & \textbf{0.857\,(0.809, 0.905)} & \textbf{-0.959\,(-1.160, -0.758)} & \textbf{0.235\,(0.196, 0.274)}\\
        & HK  &  7.454\,(7.140, 7.768) & \textbf{1.172\,(1.148, 1.196)} &  4.345\,(4.019, 4.670) & \textbf{0.550\,(0.489, 0.611)}\\
        \midrule
        \multirow{5}{*}{\textbf{CIFAR-100}}
        & TOP & \textbf{-0.311\,(-0.381, -0.241)} & \textbf{1.180\,(1.142, 1.187)} & \textbf{-1.579\,(-1.745, -1.413)} & \textbf{0.381\,(0.329, 0.433)}\\
        & WD  & 14.612\,(14.040, 15.184) & \textbf{0.703\,(0.653, 0.754)} & 13.344\,(12.736, 13.951) & -0.096\,(-0.158, -0.034)\\
        & SWK & 12.578\,(12.090, 13.066) & \textbf{0.751\,(0.702, 0.801)} & 11.310\,(10.783, 11.837) & -0.048\,(-0.110, 0.013)\\
        & BD  &  2.399\,(2.221, 2.577) & \textbf{1.090\,(1.050, 1.130)} &  1.131\,(0.881, 1.381) & \textbf{0.291\,(0.231, 0.352)}\\
        & HK  &  9.327\,(8.945, 9.709) & \textbf{0.741\,(0.697, 0.786)} &  8.059\,(7.631, 8.486) & -0.058\,(-0.119, 0.003)\\
        \bottomrule
    \end{tabular}
    \caption{Fully Connected Layers}
    \label{table:cifar_fcl}
  \end{subtable}
  
\vspace{1ex}
\begin{subtable}[t]{\textwidth}
    \centering
    \begin{tabular}{l l l l l l}
    \toprule
        Dataset & Method & $\Delta$ Epoch (VL) & $\Delta$ Acc.\,(VL)\% &
                       $\Delta$ Epoch (FP) & $\Delta$ Acc.\,(FP)\%\\
        \midrule
        \multirow{5}{*}{\textbf{CIFAR-10}}
        & TOP & \textbf{-0.458\,(-0.514, -0.402)} & \textbf{0.640\,(0.589, 0.690)} & \textbf{-5.330\,(-5.493, -5.167)} & -0.477\,(-0.534, -0.419)\\
        & WD  &  8.795\,(8.380, 9.211) & \textbf{0.564\,(0.450, 0.679)} &  3.924\,(3.604, 4.244) & -0.553\,(-0.671, -0.435)\\
        & SWK &  9.536\,(9.145, 9.928) & \textbf{0.953\,(0.904, 1.002)} &  4.665\,(4.420, 4.910) & -0.164\,(-0.213, -0.114)\\
        & BD  &  2.510\,(2.366, 2.653) & \textbf{0.837\,(0.795, 0.879)} & \textbf{-2.362\,(-2.463, -2.261)} & -0.280\,(-0.322, -0.238)\\
        & HK  &  7.236\,(6.936, 7.536) & \textbf{1.124\,(1.100, 1.150)} &  2.365\,(2.216, 2.514) &  0.007\,(-0.011, 0.026)\\
        \midrule
        \multirow{5}{*}{\textbf{CIFAR-100}}
        & TOP & \textbf{-0.742\,(-0.789, -0.695)} & \textbf{0.820\,(0.751, 0.889)} & \textbf{-8.864\,(-9.192, -8.535)} & -0.096\,(-0.160, -0.032)\\
        & WD  &  6.689\,(6.444, 6.934) & \textbf{0.478\,(0.383, 0.573)} & \textbf{-1.433\,(-1.641, -1.225)} & -0.438\,(-0.519, -0.357)\\
        & SWK &  7.480\,(7.209, 7.751) & \textbf{0.779\,(0.731, 0.828)} & \textbf{-0.642\,(-0.759, -0.525)} & -0.137\,(-0.162, -0.113)\\
        & BD  &  2.244\,(2.123, 2.365) & \textbf{0.772\,(0.704, 0.839)} & \textbf{-5.877\,(-6.128, -5.627)} & -0.144\,(-0.200, -0.088)\\
        & HK  &  5.452\,(5.261, 5.643) & \textbf{0.942\,(0.904, 0.980)} & \textbf{-2.670\,(-2.843, -2.497)} & \textbf{0.026\,(0.009, 0.042)}\\
        \bottomrule
    \end{tabular}
    \caption{Convolution Layers}
    \label{table:cifar_cl}
  \end{subtable}
  \caption{Differences in training performance between topological time series and baselines on CIFAR-10 and CIFAR-100. Results are reported in mean and 95\% CI. Earlier stops (negative gap) and higher test accuracy (positive gap) are \textbf{bolded}.}
  \label{table:cifar}
\end{table*}

\subsection*{Scenario 2: Architecture-Limited Regime}

In the architecture-limited regime, we assess early stopping criteria in settings where model capacity is constrained relative to dataset complexity. All topological time series methods, TOP, WD, BD, HK, and SWK, as well as the state-of-the-art FP baseline, are applied to models trained on the full training set. For the VL baseline, since CIFAR-10 and CIFAR-100 provide only training and test sets (with no dedicated validation set), we construct a validation set by splitting the training set: 90\% of the data is used for training and 10\% is held out for validation. We use the VGG architecture on both CIFAR-10 and CIFAR-100. Under these conditions, the test accuracy, measured at the maximum training epoch, is $75.09\% \pm 0.79\%$ for CIFAR-10 and $34.86\% \pm 1.18\%$ for CIFAR-100 (noting that CIFAR-100 contains 100 classes, so random guessing would achieve only $1\%$ accuracy).

We evaluate topological time series criteria separately on fully connected layers and convolutional layers, motivated by the distinct functional roles and activation patterns of these two types of layers within the network.

Performance differences for the fully connected and convolutional layers are summarized in Tables~\ref{table:cifar_fcl} and~\ref{table:cifar_cl}. On average, our TOP method achieves both higher test accuracy (0.954\% and 1.180\%) and earlier stopping ($-0.413$ and $-0.311$ epochs) compared to VL on CIFAR-10 and CIFAR-100, respectively. TOP also outperforms the state-of-the-art FP approach, yielding higher test accuracy (0.332\% and 0.381\%) and earlier stops in training ($-3.522$ and $-1.579$ epochs) on CIFAR-10 and CIFAR-100. These improvements are statistically significant, as the 95\% confidence intervals for the test accuracy differences are well above zero and those for the early stopping epoch differences are well below zero, indicating that TOP better captures the training dynamics. For convolutional layers, applying the early stopping criterion with TOP similarly leads to both earlier stopping and higher test accuracy than VL, while compared to FP, TOP stops earlier but with slightly lower test accuracy. In general, other methods tend to achieve higher test accuracy only at the cost of longer training times compared to the baselines.

We note that the parameter $\epsilon$ in the early stopping criterion controls sensitivity to noise in the topological time series and can influence stopping behavior. Results for additional values of $\epsilon$ are provided in the Appendix.

\begin{table}[t]
  \centering
  \customsmall
  \setlength{\tabcolsep}{1mm}
  \begin{tabular}{l c c c}
    \toprule
    Method & Time/Epoch (s) & GPU Mem (MB) & CPU Mem (MB) \\
    \midrule
    TOP  & $1.223\pm0.074$ & 54.0 & 907.2 \\
    FP   & $1.381\pm0.101$ & 54.0 & 920.5 \\
    WD   & $1.424\pm0.090$ & 54.0 & 921.5 \\
    SWK  & $1.471\pm0.080$ & 54.0 & 921.3 \\
    VL   & $1.578\pm0.084$ & 54.0 & 859.7 \\
    BD   & $3.280\pm0.650$ & 54.0 & 920.8 \\
    HK   & $3.423\pm0.536$ & 54.0 & 924.2 \\
    \bottomrule
  \end{tabular}
  \caption{Comparison of methods by time per epoch and maximum memory usage. Time per epoch is reported as mean ± standard deviation over 50 training epochs.}
  \label{tab:methods-performance}
\end{table}

\paragraph{Runtime study.}

Our proposed method (TOP) runs faster per epoch than both the VL method and the state-of-the-art FP, as shown in Table~\ref{tab:methods-performance}. TOP achieves a $22.5\%$ reduction in time per epoch compared to VL and is $11.4\%$ faster than FP. In addition to its speed, TOP also requires less CPU memory than FP and only a modest $5.5\%$ increase relative to VL. The VL method is slower primarily due to additional data transfer of the validation set (1\% of the training data) between GPU and CPU during each epoch, which incurs I/O overhead. When validation data is preloaded onto the GPU, the average epoch time for VL is reduced to $1.051 \pm 0.062$ seconds. Hardware details for these experiments are provided in the Appendix.

\section*{Limitations}
Extracting loops using persistent graph homology and persistent homology has runtimes of $\mathcal{O}(n^2 \log n)$ and $\mathcal{O}(n^3)$, respectively, where $n$ is the number of neurons in the functional connectome. For large networks, these computations can become intensive; however, it is generally unnecessary to compute large connectomes for the entire architecture, as targeted analysis of specific layers tends to yield more interpretable and effective topological insights. Indeed, our runtime experiments demonstrate that with 400 neurons, both methods not only run faster than validation loss calculations, due to fewer I/O operations during training, but also achieve higher performance than validation loss.

\bibliographystyle{plainnat}
\bibliography{reference}

\begin{thebibliography}{23}
\providecommand{\natexlab}[1]{#1}
\providecommand{\url}[1]{\texttt{#1}}
\expandafter\ifx\csname urlstyle\endcsname\relax
  \providecommand{\doi}[1]{doi: #1}\else
  \providecommand{\doi}{doi: \begingroup \urlstyle{rm}\Url}\fi

\bibitem[Carriere et~al.(2017)Carriere, Cuturi, and Oudot]{carriere2017sliced}
Mathieu Carriere, Marco Cuturi, and Steve Oudot.
\newblock Sliced wasserstein kernel for persistence diagrams.
\newblock In \emph{International Conference on Machine Learning}, pages 664--673. PMLR, 2017.

\bibitem[Edelsbrunner and Harer(2022)]{edelsbrunner2022computational}
Herbert Edelsbrunner and John~L Harer.
\newblock \emph{Computational Topology: An Introduction}.
\newblock American Mathematical Society, 2022.

\bibitem[Goodfellow et~al.(2016)Goodfellow, Bengio, and Courville]{goodfellow2016deep}
Ian Goodfellow, Yoshua Bengio, and Aaron Courville.
\newblock \emph{Deep Learning}.
\newblock MIT Press, 2016.

\bibitem[He et~al.(2016)He, Zhang, Ren, and Sun]{he2016resnet}
Kaiming He, Xiangyu Zhang, Shaoqing Ren, and Jian Sun.
\newblock Deep residual learning for image recognition.
\newblock In \emph{Proceedings of the IEEE Conference on Computer Vision and Pattern Recognition}, pages 770--778, 2016.

\bibitem[Hofer et~al.(2017)Hofer, Kwitt, Niethammer, and Uhl]{hofer2017deep}
Christoph Hofer, Roland Kwitt, Marc Niethammer, and Andreas Uhl.
\newblock Deep learning with topological signatures.
\newblock \emph{Advances in Neural Information Processing Systems}, 30, 2017.

\bibitem[Kingma and Ba(2015)]{KingmaB14adam}
Diederik~P. Kingma and Jimmy Ba.
\newblock Adam: A method for stochastic optimization.
\newblock \emph{International Conference on Learning Representations}, 2015.

\bibitem[Krizhevsky et~al.(2009)Krizhevsky, Hinton, et~al.]{krizhevsky2009learning}
Alex Krizhevsky, Geoffrey Hinton, et~al.
\newblock Learning multiple layers of features from tiny images.
\newblock 2009.

\bibitem[LeCun et~al.(1998)LeCun, Bottou, Bengio, and Haffner]{lecun1998gradient}
Yann LeCun, L{\'e}on Bottou, Yoshua Bengio, and Patrick Haffner.
\newblock Gradient-based learning applied to document recognition.
\newblock \emph{Proceedings of the IEEE}, 86\penalty0 (11):\penalty0 2278--2324, 1998.

\bibitem[Leonardi and Van De~Ville(2015)]{leonardi2015spurious}
Nora Leonardi and Dimitri Van De~Ville.
\newblock On spurious and real fluctuations of dynamic functional connectivity during rest.
\newblock \emph{NeuroImage}, 104:\penalty0 430--436, 2015.

\bibitem[Maas et~al.(2013)Maas, Hannun, Ng, et~al.]{maas2013rectifier}
Andrew~L Maas, Awni~Y Hannun, Andrew~Y Ng, et~al.
\newblock Rectifier nonlinearities improve neural network acoustic models.
\newblock In \emph{Proc. icml}, volume~30, page~3. Atlanta, GA, 2013.

\bibitem[Prechelt(1998)]{prechelt1998automatic}
Lutz Prechelt.
\newblock Automatic early stopping using cross validation: Quantifying the criteria.
\newblock \emph{Neural Networks}, 11\penalty0 (4):\penalty0 761--767, 1998.

\bibitem[Recht et~al.(2019)Recht, Roelofs, Schmidt, and Shankar]{recht2019imagenet}
Benjamin Recht, Rebecca Roelofs, Ludwig Schmidt, and Vaishaal Shankar.
\newblock Do {ImageNet} classifiers generalize to {ImageNet}?
\newblock In \emph{International Conference on Machine Learning}, pages 5389--5400, 2019.

\bibitem[Reininghaus et~al.(2015)Reininghaus, Huber, Bauer, and Kwitt]{reininghaus2015stable}
Jan Reininghaus, Stefan Huber, Ulrich Bauer, and Roland Kwitt.
\newblock A stable multi-scale kernel for topological machine learning.
\newblock In \emph{Proceedings of the IEEE Conference on Computer Vision and Pattern Recognition}, pages 4741--4748, 2015.

\bibitem[Rieck et~al.(2019)Rieck, Togninalli, Bock, Moor, Horn, Gumbsch, and Borgwardt]{rieck2018neural}
Bastian Rieck, Matteo Togninalli, Christian Bock, Michael Moor, Max Horn, Thomas Gumbsch, and Karsten Borgwardt.
\newblock Neural persistence: A complexity measure for deep neural networks using algebraic topology.
\newblock In \emph{International Conference on Learning Representations}, 2019.

\bibitem[Simonyan and Zisserman(2014)]{simonyan2014very}
Karen Simonyan and Andrew Zisserman.
\newblock Very deep convolutional networks for large-scale image recognition.
\newblock \emph{arXiv preprint arXiv:1409.1556}, 2014.

\bibitem[Skraba and Turner(2020)]{skraba2020wasserstein}
Primo{\v z} Skraba and Katharine Turner.
\newblock Wasserstein stability for persistence diagrams.
\newblock \emph{arXiv preprint arXiv:2006.16824}, 2020.

\bibitem[Songdechakraiwut et~al.(2023)Songdechakraiwut, Krause, Banks, Nourski, and Van~Veen]{songdechakraiwut2023wasserstein}
T.~Songdechakraiwut, B.~M. Krause, M.~I. Banks, K.~V. Nourski, and B.~D. Van~Veen.
\newblock Wasserstein distance-preserving vector space of persistent homology.
\newblock In \emph{Proceedings of the International Conference on Medical Image Computing and Computer-Assisted Intervention (MICCAI)}, pages 277--286, 2023.

\bibitem[Songdechakraiwut and Chung(2023)]{songdechakraiwut2023topological}
Tananun Songdechakraiwut and Moo~K Chung.
\newblock Topological learning for brain networks.
\newblock \emph{The Annals of Applied Statistics}, 17\penalty0 (1):\penalty0 403--433, 2023.

\bibitem[Songdechakraiwut and Wu(2025)]{songdechakraiwut2025functional}
Tananun Songdechakraiwut and Yutong Wu.
\newblock Functional connectomes of neural networks.
\newblock In \emph{Proceedings of the AAAI Conference on Artificial Intelligence}, volume~39, pages 20558--20566, 2025.

\bibitem[Songdechakraiwut et~al.(2021)Songdechakraiwut, Shen, and Chung]{songdechakraiwut2021topological}
Tananun Songdechakraiwut, Li~Shen, and Moo Chung.
\newblock Topological learning and its application to multimodal brain network integration.
\newblock In \emph{Proceedings of the International Conference on Medical Image Computing and Computer-Assisted Intervention (MICCAI)}, pages 166--176, 2021.

\bibitem[Vaswani et~al.(2017)Vaswani, Shazeer, Parmar, Uszkoreit, Jones, Gomez, Kaiser, and Polosukhin]{vaswani2017attention}
Ashish Vaswani, Noam Shazeer, Niki Parmar, Jakob Uszkoreit, Llion Jones, Aidan~N. Gomez, Lukasz Kaiser, and Illia Polosukhin.
\newblock Attention is all you need.
\newblock In \emph{Advances in Neural Information Processing Systems}, volume~30, 2017.

\bibitem[Xiao et~al.(2017)Xiao, Rasul, and Vollgraf]{xiao2017fashion}
Han Xiao, Kashif Rasul, and Roland Vollgraf.
\newblock Fashion-mnist: a novel image dataset for benchmarking machine learning algorithms.
\newblock \emph{arXiv preprint arXiv:1708.07747}, 2017.

\bibitem[Zhang and Lin(2024)]{zhang2024functional}
Ben Zhang and Hongwei Lin.
\newblock Functional loops: Monitoring functional organization of deep neural networks using algebraic topology.
\newblock \emph{Neural Networks}, 174, 2024.

\end{thebibliography}

\clearpage
\appendix

\section{Related Work}
Topology has become an increasingly valuable tool for understanding deep neural networks. Many recent studies have applied persistent homology to improve interpretability in deep learning models. In particular, persistent homology has emerged as a rigorous framework for capturing multiscale topological features in neural activations and connectivity patterns~\citep{rieck2018neural, hofer2017deep, zhang2024functional, songdechakraiwut2025functional}. Recent work by \citet{zhang2024functional} represents a major advance in functional topological analysis. They introduced the concept of functional loops, which are one-dimensional homological features extracted from graphs built on neural activation correlations. Their method constructs functional graphs based on activation similarity and uses persistent homology to capture higher-order interactions between neurons. Based on those insights, the authors propose a new measure called functional persistence to quantify model complexity. They also use this measure to define an early stopping criterion that does not require a validation set. However, their approach does not explicitly model how topological structures evolve over time. As a result, there remains room for extending this work to dynamic learning analysis. 

Beyond the analysis of static topological snapshots, recent studies have begun to examine how topological structures evolve throughout training. Graph filtration methods provide an efficient way to construct multiscale graph representations without relying on arbitrary thresholding~\cite{songdechakraiwut2025functional}. Building on this, a common approach is to compute the Wasserstein distance between topological summaries (e.g., persistence vectors) across consecutive training epochs. This produces a topological time series that captures the magnitude and direction of structural change over time. Such sequences have been used to detect phase transitions and measure model stability~\cite{skraba2020wasserstein}. However, most of these studies focus on visualization or post-hoc analysis, and few have integrated topological dynamics directly into the training process or optimization objectives.

\section{Persistence-Based Topological Time Series}
\subsection{Bottleneck Distance (BD)}\label{sec:BD}

The bottleneck distance (BD) quantifies the maximum discrepancy between two persistence diagrams.  
Each persistence diagram \( D \) is a multiset of points \( (b, d) \), where \( b \) and \( d \) denote the birth and death times of topological features.

To compare two diagrams \( D_1 \) and \( D_2 \), we consider all valid matchings \( \gamma \) that assign each point in \( D_1 \cup D_2 \) to exactly one of the following:
\begin{itemize}
    \item a point in the other diagram, or
    \item a point on the diagonal \( \Delta = \{(x, x) \mid x \in \mathbb{R} \} \),
\end{itemize}
such that each point is matched exactly once and no two points are matched to the same image (i.e., \( \gamma \) defines a partial bijection between \( D_1 \cup D_2 \) and itself extended with diagonal matches).  
Unmatched points in either diagram are paired with their closest diagonal projection.

The bottleneck distance is then defined as:
\[
d_B(D_1, D_2) = \inf_{\gamma} \sup_{p \in D_1 \cup D_2} \| p - \gamma(p) \|_\infty
\]
where:
\begin{itemize}
    \item If \( \gamma(p) \in D_1 \cup D_2 \), then \( \|p - \gamma(p)\|_\infty = \max \{ |b_p - b_{\gamma(p)}|, |d_p - d_{\gamma(p)}| \} \).
    \item If \( \gamma(p) \in \Delta \), then the distance is:
    \[
    \|p - \gamma(p)\|_\infty = \frac{|d_p - b_p|}{2},
    \]
    which is the minimum distance from \( p \) to the diagonal under the \( \ell_\infty \)-norm.
\end{itemize}

This metric captures the largest single topological change required to transform one diagram into the other.  
It is widely used to assess the stability of persistent homology and to compare topological features across spaces.

\subsection{Heat Kernel (HK)} \label{HK}

The heat kernel (HK) is a similarity function defined on persistence diagrams. The kernel maps persistence diagrams into a reproducing kernel Hilbert space (RKHS), enabling the use of classical machine learning algorithms such as support vector machines and $k$-means clustering ~\cite{reininghaus2015stable}. 

Let $D_1$ and $D_2$ be two persistence diagrams. For any point $p = (b_p, d_p) \in D_1$ and $q = (b_q, d_q) \in D_2$, define their reflections over the diagonal as $\bar{p} = (d_p, b_p)$ and $\bar{q} = (d_q, b_q)$. The heat kernel is then defined as:
\begin{align*}
k_\sigma(D_1, D_2) = \frac{1}{8\pi\sigma} \sum_{\substack{p \in D_1 \\ q \in D_2}} \bigg[ 
\exp\left(-\frac{\|p - q\|^2}{8\sigma}\right) \\
- \exp\left(-\frac{\|p - \bar{q}\|^2}{8\sigma}\right) 
- \exp\left(-\frac{\|\bar{p} - q\|^2}{8\sigma}\right) \\
+ \exp\left(-\frac{\|\bar{p} - \bar{q}\|^2}{8\sigma}\right) 
\bigg]
\end{align*}

The four terms capture the interaction between original points and their diagonal reflections. This ensures the kernel is symmetric and positive definite. The subtractions reduce the influence of low-persistence features and make the kernel more robust to noise.

\subsection{Sliced Wasserstein Kernel (SWK)} \label{SWK}

The sliced Wasserstein kernel (SWK) is a positive definite kernel for persistence diagrams  ~\cite{carriere2017sliced}. It combines optimal transport with kernel methods yet stays computationally efficient.

Given two persistence diagrams $D_1$ and $D_2$, the sliced Wasserstein distance of order $p$ is defined as:
\[
\mathrm{SW}_p^p(D_1, D_2) = \frac{1}{\pi} \int_{0}^{\pi} W_p^p\left( \theta_* D_1, \theta_* D_2 \right) d\theta,
\]
where $\theta_* D$ denotes the projection of all points in $D$ onto the line through the origin at angle $\theta$ and \(W_{p}\) denotes the \(p\)-Wasserstein distance. 

The sliced Wasserstein kernel is then given by:
\[
k_{\text{SW}}(D_1, D_2) = \exp\left( - \frac{\mathrm{SW}_p^p(D_1, D_2)}{2\tau} \right),
\]
where $\tau > 0$ is a bandwidth parameter.

\section{Implementation Details for Topological Time Series} \label{Implementation}
This section explains how we compute and analyze topological time series in our experiments.
\subsection{Persistence Vector (Graph Filtration)}
At each epoch, we start with a symmetric adjacency matrix~$\mathbf{M}$, which describes pairwise relationships between neurons. We use this matrix to build a weighted, undirected graph. The weight of each edge is given by~$\mathbf{M}_{ij}$.

We extract the maximum spanning tree (MST) using NetworkX’s \texttt{maximum\_spanning\_tree()} function. The MST contains the strongest connections and ensures that the graph remains connected. After constructing the MST, we remove its edges from the graph using \texttt{G.remove\_edges\_from(T.edges())}. The remaining edges, which are not part of the MST, each form a unique cycle in the graph. We sort the weights of these cycle edges in descending order. This sorted list forms the persistence vector for that epoch. (See \texttt{adj2pers} and \texttt{pers2vec})

\subsection{Persistence Diagram (Vietoris–Rips Filtration)}
We also analyze the topology using persistence diagrams. First, we convert the correlation matrix~$\mathbf{C}$ into a dissimilarity matrix, $D_{ij} = 1 - |\mathbf{C}_{ij}|$. We use the \texttt{ripser} package with \texttt{maxdim=1} to compute 1-dimensional persistence diagrams from this matrix. We always focus on $H_1$ (loops). Each diagram records the birth and death of topological loops at different scales.

\subsection{Distance Computation and Time Series Construction}
We compare topological features across epochs using different distances:
\begin{itemize}
    \item For graph filtration, we use the 1D Wasserstein distance to compare persistence vectors between epochs. We compute this with \texttt{scipy.stats.wasserstein\_distance}, using $p=1$,
    \item For persistence diagrams, we use the \texttt{persim} library, it provides functions for the $2$-Wasserstein distance, bottleneck distance, sliced Wasserstein, and heat kernel distances, all with $p=2$ by default.
\end{itemize}
For each model, we compute the chosen topological vector or diagram at every epoch. We then measure the distance between consecutive epochs. This produces a time series that tracks how the network's topology changes over time.

\section{Summary Statistics for Paired Differences}
Given a method and a baseline evaluated over \( n \) configurations, we define the accuracy difference for each configuration as:
\[
d^{(j)} = a_{\text{method}}^{(j)} - a_{\text{baseline}}^{(j)}, \quad j = 1, 2, \dots, n
\]
The sample mean of the accuracy differences is computed as:
\[
\bar{d} = \frac{1}{n} \sum_{j=1}^{n} d^{(j)}
\]
The 95\% confidence interval for the true mean difference \( \mu_d \) is given by:
\[
\text{CI}_{95\%} = \bar{d} \pm t_{n-1}^* \cdot \frac{s_d}{\sqrt{n}}
\]
Where:
\begin{itemize}
    \item \( s_d = \sqrt{ \frac{1}{n - 1} \sum_{j=1}^{n} (d^{(j)} - \bar{d})^2 } \) is the sample standard deviation of the differences.
    \item \( t_{n-1}^* \) is the critical value from the Student's t-distribution with \( n - 1 \) degrees of freedom.
\end{itemize}

A similar computation is applied for computing the sample mean for epoch differences and the 95\% confidence interval for true mean epoch differences. For early stopping experiments, each configuration is uniquely defined by a combination of burn-in rate $b$ and patience window $p$. The number of configurations is thus $n=\frac{T(T+1)}{2}$ where $T$ is the total number of epochs under the constraint $b + p \leq T$.

\section{Experimental Setup}
\subsection*{Training}
We evaluated two neural network architectures in experiments: a 2-layer MLP for MNIST and Fashion-MNIST, and three VGG blocks followed by a 2-layer MLP for CIFAR-10 and CIFAR-100~\cite{simonyan2014very}. To prevent dying ReLU, we adopted Leaky ReLU as the activation function with slope $\alpha=0.01$ \cite{maas2013rectifier}. For MNIST and Fashion-MNIST, we used stochastic gradient descent as the optimizer with a learning rate of $0.1$ and a momentum of $0.9$. For CIFAR-10 and CIFAR-100, we used the Adam optimizer with the learning rate of $0.0003$ \cite{KingmaB14adam}. All models were trained for 50 epochs with a batch size of 32.

\subsection*{System Configuration}
All experiments were performed on the system with the following specifications:
\begin{itemize}[leftmargin=*]
  \item \textbf{Components}
    \begin{itemize}[leftmargin=1.5em]
      \item \textbf{CPU:} Intel Xeon Gold 5317, 3.0 GHz
      \item \textbf{GPU:} NVIDIA RTX A5000, 24 GB VRAM
      \item \textbf{DRAM:} 128 GB
      \item \textbf{Architecture:} x86\_64
      \item \textbf{OS:} Ubuntu 22.04.5 LTS “Jammy Jellyfish”
      \item \textbf{Kernel:} Linux 5.15
      \item \textbf{CUDA Toolkit:} 12.8
      \item \textbf{NVIDIA Driver:} 570.133.20
    \end{itemize}

  \item \textbf{Key Python Libraries}
    \begin{itemize}[leftmargin=*]
      \item \textbf{Python}: 3.11.8
      \item \textbf{NumPy}: 1.26.4
      \item \textbf{Pandas}: 2.2.2
      \item \textbf{NetworkX}: 3.2.1
      \item \textbf{SciPy}: 1.13.1
      \item \textbf{PyTorch}: 2.3.1
      \item \textbf{Torchvision}: 0.18.1
      \item \textbf{scikit--learn}: 1.3.2
      \item \textbf{Ripser}: 0.6.8
      \item \textbf{Persim}: 0.3.5
    \end{itemize}
  
\end{itemize}

\section{Early Stopping Sensitivity}
The early stopping criterion proposed in the main text is parameterized by the burn-in rate $b$, the patience window $p$, and a sensitivity threshold $\epsilon$. The burn-in rate $b$ specifies the number of initial epochs excluded from evaluation, the patience window $p$ specifies the number of consecutive epochs without improvement that are allowed, and the sensitivity threshold $\epsilon$ controls the sensitivity to noise in the series. While it is possible to sweep over all $(b,p)$ configurations satisfying $b + p \leq T$, where $T$ is the number of training epochs, the choice of $\epsilon$ is arbitrary. The experimental results in the main text use $\epsilon=0.01$. The additional results for $\epsilon=0.0001$, $\epsilon=0.001$ and $\epsilon=0.1$ are presented in Tables \ref{table:mnist55_additional}, \ref{table:mnist91_additional}, \ref{table:cifar_fcn_additional} and \ref{table:cifar_cnn_additional}. Table \ref{table:mnist55_additional} and Table \ref{table:mnist91_additional} compile the additional results for the Data-Limited Regime on MNIST and Fashion-MNIST (Scenario 1), and Table \ref{table:cifar_fcn_additional} and Table \ref{table:cifar_cnn_additional} compile the additional results for the Architecture-Limited Regime on CIFAR-10/CIAFR-100 (Scenario 2). 

\begin{table*}
  \centering
  \customsmall
  \setlength{\tabcolsep}{2.7mm}
  

  \begin{subtable}[t]{\textwidth}
    \centering
    \begin{tabular}{l l l l l l}
    \toprule
      Dataset & Method & $\Delta$ Epoch (VL) & $\Delta$ Acc.\,(VL)\% &
                       $\Delta$ Epoch (FP) & $\Delta$ Acc.\,(FP)\%\\
      \midrule
        \multirow{5}{*}{\textbf{MNIST}}
        & TOP & 5.219\,(\,4.909, 5.529\,) & \textbf{2.483\,(\,2.470, 2.496\,)} & \textbf{-5.832\,(\,-6.000, -5.664\,)} & -0.023\,(\,-0.035, -0.011\,)\\
        & WD  & 20.803\,(\,20.237, 21.368\,) & \textbf{2.545\,(\,2.528, 2.562\,)} & 9.752\,(\,9.429, 10.075\,) & \textbf{0.038\,(\,0.022, 0.054\,)}\\
        & SWK & 20.929\,(\,20.361, 21.497\,) & \textbf{2.564\,(\,2.551, 2.577\,)} & 9.878\,(\,9.554, 10.202\,) & \textbf{0.057\,(\,0.046, 0.069\,)}\\
        & BD  & 19.408\,(\,18.823, 19.993\,) & \textbf{2.545\,(\,2.529, 2.560\,)} & 8.357\,(\,8.022, 8.692\,) & \textbf{0.038\,(\,0.024, 0.053\,)}\\
        & HK  & 11.701\,(\,11.263, 12.140\,) & \textbf{2.488\,(\,2.476, 2.500\,)} & 0.651\,(\,0.435, 0.866\,) & -0.018\,(\,-0.029, -0.007\,)\\
      \midrule
        \multirow{5}{*}{\shortstack{\textbf{Fashion-}\\\textbf{MNIST}}}
        & TOP & 6.517\,(\,6.209, 6.825\,) & \textbf{1.663\,(\,1.643, 1.683\,)} & \textbf{-8.051\,(\,-8.346, -7.755\,)} & -0.257\,(\,-0.276, -0.239\,)\\
        & WD  & \textbf{-1.280\,(\,-1.592, -0.968\,)} & \textbf{0.676\,(\,0.587, 0.766\,)} & \textbf{-15.847\,(\,-16.553, -15.141\,)} & -1.244\,(\,-1.337, -1.151\,)\\
        & SWK & 0.598\,(\,0.291, 0.905\,) & \textbf{0.957\,(\,0.892, 1.022\,)} & \textbf{-13.969\,(\,-14.622, -13.317\,)} & -0.963\,(\,-1.031, -0.895\,)\\
        & BD  & 6.768\,(\,6.434, 7.103\,) & \textbf{1.489\,(\,1.463, 1.515\,)} & \textbf{-7.799\,(\,-8.149, -7.449\,)} & -0.431\,(\,-0.457, -0.405\,)\\
        & HK  & 5.856\,(\,5.570, 6.142\,) & \textbf{1.526\,(\,1.502, 1.551\,)} & \textbf{-8.711\,(\,-9.088, -8.335\,)} & -0.394\,(\,-0.419, -0.369\,)\\
    \bottomrule
    \end{tabular}
    \caption{5\% Train / 5\% Validation, $\epsilon=0.0001$}
    \label{table:mnist5to5e00001}
    \end{subtable}

    \vspace{1ex}
    
      \begin{subtable}[t]{\textwidth}
        \centering
        \begin{tabular}{l l l l l l}
        \toprule
          Dataset & Method & $\Delta$ Epoch (VL) & $\Delta$ Acc.\,(VL)\% &
                           $\Delta$ Epoch (FP) & $\Delta$ Acc.\,(FP)\%\\
          \midrule
            \multirow{5}{*}{\textbf{MNIST}}
            & TOP & \textbf{-2.131\,(-2.218, -2.043)} & \textbf{2.191\,(2.174, 2.208)} & \textbf{-12.328\,(-12.639, -12.016)} & -0.321\,(-0.342, -0.300)\\
            & WD  & 20.957\,(20.389, 21.524) & \textbf{2.559\,(2.542, 2.576)} & 10.759\,(10.435, 11.084) & \textbf{0.047\,(0.031, 0.063)}\\
            & SWK & 20.951\,(20.385, 21.516) & \textbf{2.577\,(2.563, 2.590)} & 10.753\,(10.430, 11.076) & \textbf{0.065\,(0.053, 0.076)}\\
            & BD  & 16.175\,(15.619, 16.731) & \textbf{2.537\,(2.520, 2.554)} & 5.978\,(5.672, 6.283) & \textbf{0.025\,(0.010, 0.041)}\\
            & HK  &  8.820\,(8.477, 9.163) & \textbf{2.488\,(2.476, 2.499)} & \textbf{-1.377\,(-1.513, -1.242)} & -0.024\,(-0.035, -0.014)\\
          \midrule
            \multirow{5}{*}{\shortstack{\textbf{Fashion-}\\\textbf{MNIST}}}
            & TOP & \textbf{-0.224\,(-0.354, -0.094)} & \textbf{1.301\,(1.273, 1.329)} & \textbf{-14.751\,(-15.263, -14.239)} & -0.643\,(-0.675, -0.611)\\
            & WD  & \textbf{-1.096\,(-1.402, -0.789)} & \textbf{0.696\,(0.607, 0.786)} & \textbf{-15.623\,(-16.325, -14.920)} & -1.248\,(-1.341, -1.154)\\
            & SWK &  0.712\,(0.411, 1.012) & \textbf{0.973\,(0.909, 1.037)} & \textbf{-13.815\,(-14.465, -13.166)} & -0.971\,(-1.039, -0.902)\\
            & BD  &  5.287\,(4.990, 5.585) & \textbf{1.426\,(1.397, 1.454)} & \textbf{-9.240\,(-9.645, -8.834)} & -0.518\,(-0.549, -0.488)\\
            & HK  &  5.874\,(5.590, 6.158) & \textbf{1.554\,(1.531, 1.578)} & \textbf{-8.653\,(-9.031, -8.274)} & -0.390\,(-0.414, -0.365)\\
        \bottomrule
    \end{tabular}
    \caption{5\% Train / 5\% Validation, $\epsilon=0.001$}
    \label{table:mnist5to5e0001}
  \end{subtable}

  \vspace{1ex}
  
  \begin{subtable}[t]{\textwidth}
    \centering
    \begin{tabular}{l l l l l l}
    \toprule
      Dataset & Method & $\Delta$ Epoch (VL) & $\Delta$ Acc.\,(VL)\% &
                       $\Delta$ Epoch (FP) & $\Delta$ Acc.\,(FP)\%\\
      \midrule
        \multirow{5}{*}{\textbf{MNIST}}
        & TOP & \textbf{-0.914\,(-1.022, -0.806)} & \textbf{2.089\,(1.983, 2.194)} & \textbf{-2.227\,(-2.417, -2.037)} & -0.875\,(-0.993, -0.758)\\
        & WD  & 21.500\,(20.803, 22.196) & \textbf{3.313\,(3.263, 3.363)} & 20.186\,(19.559, 20.813) & \textbf{0.349\,(0.328, 0.370)}\\
        & SWK & 16.928\,(16.267, 17.589) & \textbf{3.306\,(3.256, 3.356)} & 15.615\,(15.029, 16.201) & \textbf{0.342\,(0.323, 0.361)}\\
        & BD  & \textbf{-0.914\,(-1.022, -0.806)} & \textbf{2.089\,(1.983, 2.194)} & \textbf{-2.227\,(-2.417, -2.037)} & -0.875\,(-0.993, -0.758)\\
        & HK  &  0.391\,(0.350, 0.432) & \textbf{2.763\,(2.739, 2.787)} & \textbf{-0.923\,(-1.018, -0.828)} & -0.201\,(-0.222, -0.180)\\
      \midrule
        \multirow{5}{*}{\shortstack{\textbf{Fashion-}\\\textbf{MNIST}}}
        & TOP & \textbf{-1.106\,(-1.223, -0.990)} & \textbf{1.441\,(1.324, 1.558)} & \textbf{-7.775\,(-8.213, -7.338)} & -1.241\,(-1.371, -1.111)\\
        & WD  &  1.198\,(1.027, 1.369) & \textbf{1.761\,(1.675, 1.846)} & \textbf{-5.471\,(-5.939, -5.004)} & -0.923\,(-1.020, -0.823)\\
        & SWK &  1.491\,(1.336, 1.646) & \textbf{1.924\,(1.859, 1.990)} & \textbf{-5.178\,(-5.620, -4.736)} & -0.759\,(-0.837, -0.680)\\
        & BD  & \textbf{-1.106\,(-1.223, -0.990)} & \textbf{1.441\,(1.324, 1.558)} & \textbf{-7.775\,(-8.213, -7.338)} & -1.241\,(-1.371, -1.111)\\
        & HK  &  1.076\,(0.984, 1.169) & \textbf{2.066\,(2.029, 2.103)} & \textbf{-5.593\,(-5.979, -5.206)} & -0.616\,(-0.669, -0.564)\\
        \bottomrule
    \end{tabular}
    \caption{5\% Train / 5\% Validation, $\epsilon=0.1$}
    \label{table:mnist5to5e01}
  \end{subtable}

\caption{Differences in training performance between topological time series and baselines on MNIST and Fashion-MNIST. 5\% of the training data is used for training, and 5\% of the training set is used as a validation set for validation loss (VL). Results are reported in sample mean and 95\% CI. Earlier stops (negative gap) and higher test accuracy (positive gap) are \textbf{bolded}.}
   \label{table:mnist55_additional}
\end{table*}

\begin{table*}
  \centering
  \customsmall
  \setlength{\tabcolsep}{2.8mm}
    \begin{subtable}[t]{\textwidth}
    \centering
    \begin{tabular}{l l l l l l}
    \toprule
      Dataset & Method & $\Delta$ Epoch (VL) & $\Delta$ Acc.\,(VL)\% &
                       $\Delta$ Epoch (FP) & $\Delta$ Acc.\,(FP)\%\\
      \midrule
        \multirow{5}{*}{\textbf{MNIST}}
        & TOP & 4.576\,(4.275, 4.876) & \textbf{0.246\,(0.232, 0.259)} & \textbf{-5.826\,(-5.989, -5.662)} & -0.025\,(-0.036, -0.013)\\
        & WD  & 20.179\,(19.620, 20.739) & \textbf{0.308\,(0.290, 0.326)} & 9.778\,(9.456, 10.100) & \textbf{0.038\,(0.022, 0.053)}\\
        & SWK & 20.262\,(19.700, 20.823) & \textbf{0.327\,(0.314, 0.340)} & 9.861\,(9.537, 10.184) & \textbf{0.056\,(0.045, 0.067)}\\
        & BD  & 18.633\,(18.059, 19.207) & \textbf{0.307\,(0.291, 0.323)} & 8.232\,(7.902, 8.562) & \textbf{0.037\,(0.023, 0.051)}\\
        & HK  & 10.838\,(10.412, 11.264) & \textbf{0.252\,(0.239, 0.264)} & 0.437\,(0.229, 0.645) & -0.019\,(-0.029, -0.008)\\
      \midrule
        \multirow{5}{*}{\shortstack{\textbf{Fashion-}\\\textbf{MNIST}}}
        & TOP & 5.736\,(5.445, 6.026) & -0.126\,(-0.147, -0.104) & \textbf{-7.969\,(-8.255, -7.684)} & -0.226\,(-0.246, -0.206)\\
        & WD  & \textbf{-2.098\,(-2.410, -1.785)} & -1.113\,(-1.205, -1.021) & \textbf{-15.803\,(-16.503, -15.103)} & -1.213\,(-1.308, -1.117)\\
        & SWK & \textbf{-0.382\,(-0.681, -0.082)} & -0.830\,(-0.895, -0.765) & \textbf{-14.087\,(-14.734, -13.440)} & -0.930\,(-0.999, -0.861)\\
        & BD  & 6.042\,(5.713, 6.371) & -0.366\,(-0.394, -0.338) & \textbf{-7.663\,(-8.001, -7.325)} & -0.466\,(-0.495, -0.438)\\
        & HK  & 5.227\,(4.959, 5.496) & -0.209\,(-0.232, -0.185) & \textbf{-8.478\,(-8.825, -8.130)} & -0.309\,(-0.332, -0.286)\\
    \bottomrule
    \end{tabular}
    \caption{9\% Train / 1\% Validation, $\epsilon=0.0001$}
    \label{table:mnist9to1e00001}
    \end{subtable}
\vspace{1ex}
    
    \begin{subtable}[t]{\textwidth}
    \centering
    \begin{tabular}{l l l l l l}
    \toprule
      Dataset & Method & $\Delta$ Epoch (VL) & $\Delta$ Acc.\,(VL)\% &
                       $\Delta$ Epoch (FP) & $\Delta$ Acc.\,(FP)\%\\
      \midrule
        \multirow{5}{*}{\textbf{MNIST}}
        & TOP & \textbf{-2.431\,(-2.530, -2.332)} & -0.043\,(-0.063, -0.022) & \textbf{-12.179\,(-12.485, -11.873)} & -0.321\,(-0.342, -0.300)\\
        & WD  & 20.638\,(20.074, 21.201) & \textbf{0.325\,(0.307, 0.343)} & 10.890\,(10.563, 11.217) & \textbf{0.047\,(0.031, 0.063)}\\
        & SWK & 20.613\,(20.052, 21.175) & \textbf{0.343\,(0.330, 0.356)} & 10.866\,(10.540, 11.192) & \textbf{0.065\,(0.054, 0.076)}\\
        & BD  & 15.715\,(15.168, 16.261) & \textbf{0.301\,(0.284, 0.319)} & 5.967\,(5.663, 6.270) & \textbf{0.023\,(0.008, 0.039)}\\
        & HK  &  8.401\,(8.064, 8.737) & \textbf{0.254\,(0.241, 0.266)} & \textbf{-1.347\,(-1.485, -1.209)} & -0.024\,(-0.035, -0.014)\\
      \midrule
        \multirow{5}{*}{\shortstack{\textbf{Fashion-}\\\textbf{MNIST}}}
        & TOP & \textbf{-1.142\,(-1.264, -1.020)} & -0.498\,(-0.527, -0.470) & \textbf{-14.807\,(-15.316, -14.299)} & -0.617\,(-0.649, -0.585)\\
        & WD  & \textbf{-1.924\,(-2.233, -1.615)} & -1.104\,(-1.196, -1.012) & \textbf{-15.590\,(-16.287, -14.893)} & -1.222\,(-1.318, -1.127)\\
        & SWK & \textbf{-0.263\,(-0.558, 0.031)} & -0.831\,(-0.897, -0.766) & \textbf{-13.929\,(-14.573, -13.285)} & -0.950\,(-1.019, -0.881)\\
        & BD  &  4.693\,(4.395, 4.992) & -0.441\,(-0.472, -0.411) & \textbf{-8.972\,(-9.361, -8.584)} & -0.560\,(-0.591, -0.528)\\
        & HK  &  5.264\,(4.996, 5.532) & -0.202\,(-0.225, -0.179) & \textbf{-8.401\,(-8.749, -8.054)} & -0.320\,(-0.344, -0.297)\\
    \bottomrule
    \end{tabular}
    \caption{9\% Train / 1\% Validation, $\epsilon=0.001$}
    \label{table:mnist9to1e0001}
\end{subtable}

\vspace{1ex}

\begin{subtable}[t]{\textwidth}
    \centering
    \begin{tabular}{l l l l l l}
    \toprule
      Dataset & Method & $\Delta$ Epoch (VL) & $\Delta$ Acc.\,(VL)\% &
                       $\Delta$ Epoch (FP) & $\Delta$ Acc.\,(FP)\%\\
      \midrule
        \multirow{5}{*}{\textbf{MNIST}}
        & TOP & \textbf{-0.806\,(-0.908, -0.703)} & -0.385\,(-0.492, -0.278) & \textbf{-2.236\,(-2.426, -2.045)} & -0.875\,(-0.991, -0.758)\\
        & WD  & 21.607\,(20.906, 22.308) & \textbf{0.837\,(0.799, 0.876)} & 20.177\,(19.552, 20.803) & \textbf{0.348\,(0.326, 0.369)}\\
        & SWK & 17.125\,(16.455, 17.796) & \textbf{0.834\,(0.795, 0.872)} & 15.696\,(15.106, 16.285) & \textbf{0.344\,(0.325, 0.363)}\\
        & BD  & \textbf{-0.806\,(-0.908, -0.703)} & -0.385\,(-0.492, -0.278) & \textbf{-2.236\,(-2.426, -2.045)} & -0.875\,(-0.991, -0.758)\\
        & HK  & 0.526\,(0.478, 0.574) & \textbf{0.297\,(0.282, 0.311)} & \textbf{-0.903\,(-0.995, -0.811)} & -0.193\,(-0.213, -0.173)\\
      \midrule
        \multirow{5}{*}{\shortstack{\textbf{Fashion-}\\\textbf{MNIST}}}
        & TOP & \textbf{-0.928\,(-1.033, -0.824)} & -0.466\,(-0.582, -0.350) & \textbf{-7.891\,(-8.335, -7.448)} & -1.267\,(-1.398, -1.136)\\
        & WD  & 1.317\,(1.160, 1.475) & -0.133\,(-0.218, -0.048) & \textbf{-5.646\,(-6.117, -5.174)} & -0.933\,(-1.033, -0.833)\\
        & SWK & 1.573\,(1.430, 1.716) & \textbf{0.018\,(-0.049, 0.085)} & \textbf{-5.390\,(-5.839, -4.941)} & -0.783\,(-0.865, -0.700)\\
        & BD  & \textbf{-0.928\,(-1.033, -0.824)} & -0.466\,(-0.582, -0.350) & \textbf{-7.891\,(-8.335, -7.448)} & -1.267\,(-1.398, -1.136)\\
        & HK  & 1.230\,(1.152, 1.309) & \textbf{0.178\,(0.144, 0.213)} & \textbf{-5.732\,(-6.121, -5.344)} & -0.622\,(-0.675, -0.570)\\
        \bottomrule
    \end{tabular}
    \caption{9\% Train / 1\% Validation, $\epsilon=0.1$}
    \label{table:mnist9to1e01}
\end{subtable}

   \caption{Differences in training performance between topological time series and baselines on MNIST and Fashion-MNIST. 9\% of the training data is used for training, and 1\% of the training set is used as a validation set for validation loss (VL). Results are reported in sample mean and 95\% CI. Earlier stops (negative gap) and higher test accuracy (positive gap) are \textbf{bolded}.}
   \label{table:mnist91_additional}
\end{table*}

\begin{table*}[ht]
  \centering
  \customsmall
  \setlength{\tabcolsep}{2.8mm}
\begin{subtable}[t]{\textwidth}
    \centering
    \begin{tabular}{l l l l l l}
    \toprule
        Dataset & Method & $\Delta$ Epoch (VL) & $\Delta$ Acc. (VL) (\%) &
                           $\Delta$ Epoch (FP) & $\Delta$ Acc. (FP) (\%)\\
    \midrule
        \multirow{5}{*}{\textbf{CIFAR-10}}
        & TOP & 9.498\,(9.080, 9.916) & \textbf{1.083\,(1.068, 1.098)} & 4.597\,(4.155, 5.040) & \textbf{0.511\,(0.455, 0.567)}\\
        & WD  & 15.624\,(15.000, 16.248) & \textbf{1.136\,(1.111, 1.161)} & 10.723\,(10.124, 11.322) & \textbf{0.564\,(0.509, 0.620)}\\
        & SWK & 13.979\,(13.414, 14.545) & \textbf{1.075\,(1.052, 1.098)} & 9.078\,(8.531, 9.625) & \textbf{0.504\,(0.449, 0.558)}\\
        & BD  & 10.803\,(10.348, 11.259) & \textbf{1.019\,(0.995, 1.044)} & 5.902\,(5.482, 6.323) & \textbf{0.448\,(0.397, 0.498)}\\
        & HK  & 10.372\,(9.948, 10.797) & \textbf{1.021\,(0.999, 1.044)} & 5.472\,(5.049, 5.894) & \textbf{0.450\,(0.397, 0.502)}\\
    \midrule
        \multirow{5}{*}{\textbf{CIFAR-100}}
        & TOP & 8.496\,(8.125, 8.867) & \textbf{0.617\,(0.574, 0.661)} & 6.219\,(5.757, 6.682) & -0.121\,(-0.174, -0.068)\\
        & WD  & 15.063\,(14.474, 15.653) & \textbf{0.571\,(0.519, 0.623)} & 12.787\,(12.142, 13.431) & -0.167\,(-0.221, -0.113)\\
        & SWK & 13.647\,(13.126, 14.168) & \textbf{0.608\,(0.557, 0.659)} & 11.370\,(10.792, 11.948) & -0.130\,(-0.183, -0.076)\\
        & BD  & 12.022\,(11.541, 12.504) & \textbf{0.664\,(0.617, 0.711)} & 9.746\,(9.211, 10.280) & -0.074\,(-0.126, -0.021)\\
        & HK  & 11.689\,(11.222, 12.156) & \textbf{0.597\,(0.551, 0.643)} & 9.413\,(8.886, 9.939) & -0.141\,(-0.194, -0.089)\\
    \bottomrule
    \end{tabular}
    \caption{Fully Connected Layers, $\epsilon=0.0001$}
    \label{table:cifar_fc_e0001}
\end{subtable}
\vspace{1ex}

  \begin{subtable}[t]{\textwidth}
    \centering
    \begin{tabular}{l l l l l l}
    \toprule
        Dataset & Method & $\Delta$ Epoch (VL) & $\Delta$ Acc. (VL) (\%) &
                           $\Delta$ Epoch (FP) & $\Delta$ Acc. (FP) (\%)\\
        \midrule
        \multirow{5}{*}{\textbf{CIFAR-10}}
        & TOP & 2.363\,(2.169, 2.557) & \textbf{0.907\,(0.893, 0.921)} & \textbf{-2.241\,(-2.573, -1.909)} & \textbf{0.339\,(0.284, 0.393)}\\
        & WD  & 15.552\,(14.933, 16.172) & \textbf{1.137\,(1.112, 1.162)} & 10.948\,(10.356, 11.541) & \textbf{0.569\,(0.513, 0.625)}\\
        & SWK & 13.983\,(13.417, 14.548) & \textbf{1.077\,(1.054, 1.100)} & 9.379\,(8.835, 9.923) & \textbf{0.509\,(0.455, 0.564)}\\
        & BD  & 9.200\,(8.797, 9.604) & \textbf{0.999\,(0.974, 1.025)} & 4.596\,(4.225, 4.968) & \textbf{0.431\,(0.383, 0.480)}\\
        & HK  & 9.939\,(9.534, 10.343) & \textbf{1.026\,(1.003, 1.049)} & 5.335\,(4.932, 5.737) & \textbf{0.458\,(0.406, 0.510)}\\
        \midrule
        \multirow{5}{*}{\textbf{CIFAR-100}}
        & TOP & 2.433\,(2.217, 2.650) & \textbf{0.855\,(0.823, 0.888)} & 0.263\,(-0.084, 0.611) & \textbf{0.119\,(0.067, 0.171)}\\
        & WD  & 15.023\,(14.434, 15.611) & \textbf{0.575\,(0.523, 0.627)} & 12.853\,(12.211, 13.495) & -0.162\,(-0.216, -0.107)\\
        & SWK & 13.614\,(13.093, 14.136) & \textbf{0.607\,(0.557, 0.658)} & 11.444\,(10.868, 12.021) & -0.129\,(-0.183, -0.076)\\
        & BD  & 9.538\,(9.138, 9.938) & \textbf{0.695\,(0.649, 0.740)} & 7.368\,(6.908, 7.828) & -0.042\,(-0.094, 0.011)\\
        & HK  & 11.453\,(10.993, 11.914) & \textbf{0.603\,(0.558, 0.649)} & 9.283\,(8.765, 9.802) & -0.133\,(-0.185, -0.081)\\
        \bottomrule
    \end{tabular}
    \caption{Fully Connected Layers, $\epsilon=0.001$}
    \label{table:cifar_fcle0001}
    \end{subtable}

    \vspace{1ex}

    \begin{subtable}[t]{\textwidth}
    \centering
    \begin{tabular}{l l l l l l}
    \toprule
        Dataset & Method & $\Delta$ Epoch (VL) & $\Delta$ Acc. (VL) (\%) &
                           $\Delta$ Epoch (FP) & $\Delta$ Acc. (FP) (\%)\\
        \midrule
        \multirow{5}{*}{\textbf{CIFAR-10}}
        & TOP & \textbf{-0.550\,(-0.597, -0.503)} & \textbf{0.073\,(-0.087, 0.232)} & \textbf{-0.622\,(-0.658, -0.586)} & -0.403\,(-0.486, -0.320)\\
        & WD  & 11.898\,(11.387, 12.408) & \textbf{1.544\,(1.495, 1.592)} & 11.825\,(11.317, 12.333) & \textbf{1.068\,(0.950, 1.186)}\\
        & SWK &  8.610\,(8.222, 8.999) & \textbf{1.470\,(1.423, 1.516)} &  8.538\,(8.151, 8.925) & \textbf{0.994\,(0.877, 1.111)}\\
        & BD  & \textbf{-0.550\,(-0.597, -0.503)} & \textbf{0.073\,(-0.087, 0.232)} & \textbf{-0.622\,(-0.658, -0.586)} & -0.403\,(-0.486, -0.320)\\
        & HK  &  2.141\,(2.029, 2.253) & \textbf{1.131\,(1.096, 1.167)} &  2.069\,(1.941, 2.196) & \textbf{0.656\,(0.567, 0.744)}\\
        \midrule
        \multirow{5}{*}{\textbf{CIFAR-100}}
        & TOP & \textbf{-0.789\,(-0.850, -0.729)} & -0.103\,(-0.308, 0.103) & \textbf{-0.196\,(-0.222, -0.170)} & -0.694\,(-0.817, -0.570)\\
        & WD  & 11.473\,(10.998, 11.947) & \textbf{0.857\,(0.815, 0.899)} & 12.066\,(11.584, 12.548) & \textbf{0.266\,(0.165, 0.367)}\\
        & SWK &  9.625\,(9.226, 10.025) & \textbf{0.849\,(0.808, 0.891)} & 10.218\,(9.811, 10.626) & \textbf{0.258\,(0.158, 0.359)}\\
        & BD  & \textbf{-0.818\,(-0.884, -0.753)} & -0.182\,(-0.403, 0.040) & \textbf{-0.225\,(-0.257, -0.193)} & -0.773\,(-0.915, -0.631)\\
        & HK  &  3.309\,(3.150, 3.468) & \textbf{1.207\,(1.178, 1.236)} &  3.902\,(3.729, 4.075) & \textbf{0.616\,(0.510, 0.722)}\\
        \bottomrule
    \end{tabular}
    \caption{Fully Connected Layers, $\epsilon=0.1$}
    \label{table:cifar_fcle01}
    \end{subtable}
  \caption{Differences in training performance between topological time series and baselines on fully connected layers on CIFAR-10 and CIFAR-100. Results are reported in mean and 95\% CI. Earlier stops (negative gap) and higher test accuracy (positive gap) are \textbf{bolded}.}
  \label{table:cifar_fcn_additional}
\end{table*}

\begin{table*}[ht]
  \centering
  \customsmall
  \setlength{\tabcolsep}{3.2mm}
\begin{subtable}[t]{\textwidth}
    \centering
    \begin{tabular}{l l l l l l}
    \toprule
        Dataset & Method & $\Delta$ Epoch (VL) & $\Delta$ Acc. (VL) (\%) &
                           $\Delta$ Epoch (FP) & $\Delta$ Acc. (FP) (\%)\\
    \midrule
        \multirow{5}{*}{\textbf{CIFAR-10}}
        & TOP & 8.790\,(8.431, 9.150) & \textbf{1.006\,(0.990, 1.022)} & 2.234\,(2.097, 2.371) & \textbf{0.020\,(0.009, 0.031)}\\
        & WD  & 9.178\,(8.757, 9.599) & \textbf{0.427\,(0.313, 0.542)} & 2.622\,(2.336, 2.908) & -0.559\,(-0.675, -0.442)\\
        & SWK & 9.497\,(9.112, 9.883) & \textbf{0.808\,(0.757, 0.859)} & 2.941\,(2.758, 3.124) & -0.178\,(-0.229, -0.127)\\
        & BD  & 8.378\,(8.043, 8.713) & \textbf{0.941\,(0.915, 0.966)} & 1.822\,(1.700, 1.944) & -0.045\,(-0.067, -0.023)\\
        & HK  & 9.479\,(9.085, 9.874) & \textbf{0.984\,(0.958, 1.014)} & 2.923\,(2.759, 3.087) & -0.002\,(-0.022, 0.018)\\
    \midrule
        \multirow{5}{*}{\textbf{CIFAR-100}}
        & TOP & 7.638\,(7.351, 7.925) & \textbf{0.755\,(0.719, 0.791)} & \textbf{-3.249\,(-3.389, -3.108)} & \textbf{0.020\,(0.014, 0.027)}\\
        & WD  & 6.909\,(6.657, 7.162) & \textbf{0.405\,(0.313, 0.498)} & \textbf{-3.978\,(-4.242, -3.713)} & -0.330\,(-0.423, -0.263)\\
        & SWK & 8.135\,(7.845, 8.424) & \textbf{0.683\,(0.635, 0.731)} & \textbf{-2.752\,(-2.927, -2.578)} & -0.052\,(-0.081, -0.023)\\
        & BD  & 8.784\,(8.457, 9.110) & \textbf{0.695\,(0.648, 0.742)} & \textbf{-2.103\,(-2.224, -1.982)} & -0.040\,(-0.065, -0.016)\\
        & HK  & 9.143\,(8.803, 9.482) & \textbf{0.754\,(0.711, 0.796)} & \textbf{-1.744\,(-1.862, -1.626)} & \textbf{0.018\,(-0.000, 0.037)}\\
    \bottomrule
    \end{tabular}
    \caption{Convolution Layers, $\epsilon=0.0001$}
    \label{table:cifar_conv_e00001}
\end{subtable}

\vspace{1ex}

    \begin{subtable}[t]{\textwidth}
    \centering
    \begin{tabular}{l l l l l l}
    \toprule
        Dataset & Method & $\Delta$ Epoch (VL) & $\Delta$ Acc. (VL) (\%) &
                           $\Delta$ Epoch (FP) & $\Delta$ Acc. (FP) (\%)\\
        \midrule
        \multirow{5}{*}{\textbf{CIFAR-10}}
        & TOP & 3.854\,(3.689, 4.020) & \textbf{0.894\,(0.879, 0.910)} & \textbf{-2.527\,(-2.627, -2.427)} & -0.090\,(-0.103, -0.078)\\
        & WD  & 9.142\,(8.725, 9.560) & \textbf{0.428\,(0.314, 0.543)} & 2.761\,(2.477, 3.045) & -0.556\,(-0.673, -0.440)\\
        & SWK & 9.487\,(9.101, 9.872) & \textbf{0.813\,(0.762, 0.864)} & 3.106\,(2.919, 3.292) & -0.172\,(-0.222, -0.121)\\
        & BD  & 7.324\,(7.023, 7.625) & \textbf{0.925\,(0.899, 0.952)} & 0.943\,(0.833, 1.053) & -0.060\,(-0.082, -0.037)\\
        & HK  & 9.284\,(8.898, 9.671) & \textbf{0.988\,(0.963, 1.014)} & 2.903\,(2.742, 3.064) & \textbf{0.004\,(-0.016, 0.024)}\\
        \midrule
        \multirow{5}{*}{\textbf{CIFAR-100}}
        & TOP & 2.674\,(2.567, 2.781) & \textbf{0.967\,(0.941, 0.992)} & \textbf{-7.996\,(-8.309, -7.683)} & \textbf{0.220\,(0.199, 0.241)}\\
        & WD  & 6.860\,(6.609, 7.110) & \textbf{0.403\,(0.310, 0.497)} & \textbf{-3.811\,(-4.072, -3.550)} & -0.343\,(-0.423, -0.263)\\
        & SWK & 8.121\,(7.831, 8.411) & \textbf{0.682\,(0.634, 0.731)} & \textbf{-2.550\,(-2.722, -2.377)} & -0.064\,(-0.093, -0.035)\\
        & BD  & 7.551\,(7.264, 7.838) & \textbf{0.689\,(0.640, 0.737)} & \textbf{-3.120\,(-3.273, -2.967)} & -0.058\,(-0.085, -0.030)\\
        & HK  & 8.818\,(8.489, 9.147) & \textbf{0.770\,(0.728, 0.812)} & \textbf{-1.853\,(-1.973, -1.732)} & \textbf{0.024\,(0.005, 0.042)}\\
        \bottomrule
    \end{tabular}
    \caption{Convolution Layers, $\epsilon=0.001$}
    \label{table:cifar_cle0001}
    \end{subtable}

\vspace{1ex}

    \begin{subtable}[t]{\textwidth}
    \centering
    \begin{tabular}{l l l l l l}
    \toprule
        Dataset & Method & $\Delta$ Epoch (VL) & $\Delta$ Acc. (VL) (\%) &
                           $\Delta$ Epoch (FP) & $\Delta$ Acc. (FP) (\%)\\
        \midrule
        \multirow{5}{*}{\textbf{CIFAR-10}}
        & TOP & \textbf{-0.527\,(-0.570, -0.485)} & \textbf{0.140\,(-0.004, 0.284)} & \textbf{-1.655\,(-1.765, -1.545)} & -1.114\,(-1.280, -0.949)\\
        & WD  &  6.447\,(6.167, 6.727) & \textbf{0.742\,(0.636, 0.848)} &  5.319\,(5.063, 5.575) & -0.512\,(-0.633, -0.391)\\
        & SWK &  5.403\,(5.189, 5.617) & \textbf{1.051\,(0.996, 1.106)} &  4.275\,(4.103, 4.447) & -0.203\,(-0.264, -0.143)\\
        & BD  & \textbf{-0.543\,(-0.589, -0.497)} & \textbf{0.081\,(-0.077, 0.239)} & \textbf{-1.671\,(-1.784, -1.558)} & -1.174\,(-1.353, -0.995)\\
        & HK  &  1.973\,(1.911, 2.036) & \textbf{1.117\,(1.091, 1.143)} &  0.846\,(0.777, 0.914) & -0.137\,(-0.163, -0.112)\\
        \midrule
        \multirow{5}{*}{\textbf{CIFAR-100}}
        & TOP & \textbf{-0.788\,(-0.848, -0.728)} & -0.097\,(-0.301, 0.107) & \textbf{-1.769\,(-1.918, -1.620)} & -1.199\,(-1.386, -1.012)\\
        & WD  &  3.768\,(3.617, 3.918) & \textbf{0.519\,(0.415, 0.622)} &  2.787\,(2.638, 2.935) & -0.584\,(-0.670, -0.498)\\
        & SWK &  3.110\,(2.991, 3.228) & \textbf{0.909\,(0.857, 0.960)} &  2.129\,(2.036, 2.221) & -0.194\,(-0.225, -0.163)\\
        & BD  & \textbf{-0.811\,(-0.875, -0.747)} & -0.166\,(-0.384, 0.053) & \textbf{-1.792\,(-1.945, -1.640)} & -1.268\,(-1.470, -1.066)\\
        & HK  &  0.945\,(0.895, 0.994) & \textbf{0.957\,(0.911, 1.003)} & \textbf{-0.036\,(-0.142, 0.069)} & -0.146\,(-0.174, -0.117)\\
        \bottomrule
    \end{tabular}
    \caption{Convolution Layers, $\epsilon=0.1$}
    \label{table:cifar_convle01}
    \end{subtable}
  
  \caption{Differences in training performance between topological time series and baselines on convolution layers on CIFAR-10 and CIFAR-100. Results are reported in mean and 95\% CI. Earlier stops (negative gap) and higher test accuracy (positive gap) are \textbf{bolded}.}
  \label{table:cifar_cnn_additional}
\end{table*}

\end{document}